\documentclass[11pt]{article} 

\usepackage[utf8]{inputenc} 

\usepackage{geometry} 
\usepackage{graphicx} 

\usepackage{algorithmicx}
\usepackage{algorithm}
\usepackage{algpseudocode}
\usepackage{amsmath}
\usepackage{amssymb}
\usepackage{longtable}

\usepackage{booktabs} 
\usepackage{array} 
\usepackage{paralist} 
\usepackage{verbatim} 
\usepackage{amsmath}
\usepackage{amssymb}
\usepackage{subfig} 

\usepackage{fancyhdr} 
\usepackage{sectsty}
\allsectionsfont{\sffamily\mdseries\upshape} 

\usepackage[nottoc,notlof,notlot]{tocbibind} 
\usepackage[titles,subfigure]{tocloft} 

\usepackage{caption} 

\title{Building Damage-Resilient Dominating Sets in Complex Networks against Random and Targeted Attacks}
\author{F. Moln\'{a}r Jr.$^{1,2,}\footnote{E-mail: molnaf@rpi.edu}$, N. Derzsy$^{1,2}$, B. K. Szymanski$^{2,3}$, G. Korniss$^{1,2}$}

\begin{document}
\maketitle

\begin{flushleft}
$^{\bf{1}}$ Department of Physics, Rensselaer Polytechnic Institute,
110 8$^{th}$ Street, Troy, NY, 12180-3590 USA \\
$^{\bf{2}}$ Social Cognitive Networks Academic Research Center,
Rensselaer Polytechnic Institute, 110 8$^{th}$ Street, Troy, NY, 12180-3590 USA \\
$^{\bf{3}}$ Department of Computer Science,
Rensselaer Polytechnic Institute, 110 8$^{th}$ Street, Troy, NY, 12180-3590 USA \\
\end{flushleft}

\section*{Abstract}

We study the vulnerability of dominating sets against random and targeted node removals in complex networks. While small, cost-efficient dominating sets play a significant role in controllability and observability of these networks, a fixed and intact network structure is always implicitly assumed. We find that cost-efficiency of dominating sets optimized for small size alone comes at a price of being vulnerable to damage; domination in the remaining network can be severely disrupted, even if a small fraction of dominator nodes are lost. We develop two new methods for finding flexible dominating sets, allowing either adjustable overall resilience, or dominating set size, while maximizing the dominated fraction of the remaining network after the attack. We analyze the efficiency of each method on synthetic scale-free networks, as well as real complex networks.

\section*{Introduction}

Dominating sets play a critical role in complex networked systems by
providing efficient sources of influence and information dispersal,
or hubs of surveillance
\cite{domination-fundamentals,Vazquez_PRE2005,nacher-mds,our-mds}, and are applied
in social, infrastructure, and communication networks
\cite{Kelleher_1988,Wang,Eubank_2004}. Most recently, dominating sets were
employed to controllability in complex networks
\cite{nacher-control,Nacher_SFREP2013,nacher-control2,Jia_NCOMM2013}, observability of the power-grid
\cite{Yang_PRL2012}, and to finding high-impact
optimal  subsets in protein interaction networks
\cite{Wuchty_PNAS2014}. While finding the smallest, most efficient
dominating set has gained significant interest, it is also important
to understand how robust these dominating sets are against various
forms of network damage.

By definition, a dominating set is a subset of nodes in a network,
such that every node not in the dominating set is adjacent to at
least one node in this set; in other words, every node has at least
one neighbor (or itself) in the dominating set. The smallest
cardinality dominating set is the minimum dominating set (MDS),
which is of particular interest, because it provides the most
cost-efficient solution for network control, assuming a constant per-node
cost of implementing control, in fixed or slowly evolving networks.
Research has been focused on finding bounds for the size of MDS
\cite{domination-fundamentals,Cooper}, finding approximations to the
MDS \cite{Potluri_2011,Hedar_2010}, understanding its expected
scaling behavior in complex networks \cite{nacher-mds,our-mds}, and
studying the impact of assortativity on network domination
\cite{Vazquez_PRE2005,our-cds}.

Attacks on complex networks, fault tolerance, and defense strategies
against damage of nodes and edges have also gained significant
interest in network science
\cite{attack-barabasi,random-fail-Duch,attack-gallos,vulnerability-Holme}.
Networks with scale-free topologies have been found to be resilient
against random node damage, but vulnerable to targeted removal of
high degree nodes \cite{cohen-1,cohen-2,callaway}. Research has also
focused on improving the robustness of these networks against
various combinations of attacks
\cite{robustness-tanizawa,robustness-tanizawa-1,robustness-tanizawa-2},
and on studying the dynamically progressing effects of an initial
damage, such as cascading failures
\cite{cascade-hayashi,cascade-andi}.

The connectivity of the surviving network structures and the
fraction of the remaining set of nodes still dominated following
failures or attacks are both essential for sustainable network
operations and carrying out network functions. While the former
(structural integrity) has been studied in great detail over the
past two decades
\cite{attack-barabasi,random-fail-Duch,attack-gallos,vulnerability-Holme,cohen-1,cohen-2,callaway},
the latter (domination stability) has not received any attention.

We assume that the network damage is relatively small, and although
the network may become fragmented due to the loss of nodes, we
assume it remains functional. In such cases efficient domination
over the network is still important and desirable, just as it is in
undamaged networks. However, considering that most dominating set
search methods aim for the smallest possible set size (and
corresponding cost) in a fixed topology network, even a small damage
could severely disrupt the complete domination ``coverage''. Our
goal is to understand how fragile dominating sets are, how to
improve them, and ultimately to provide new methods for selecting
dominating sets with adjustable balance between resilience and cost.

In order to quantify the resilience of a dominating set against network damage, we define \textit{domination stability} as the fraction of the remaining network still dominated after some nodes have been removed from the network (and thus from the dominating set):
\begin{equation}
s(f) := \frac{|\bigcup _{j \in \mathrm{DS}} \mathrm{N^{+}}(j)|}{N (1-f)},
\label{eq-def-stab}
\end{equation}
where $\mathrm{DS}$ is a dominating set of the original (undamaged) network, $f$ is the fraction of nodes removed from the network, and $\mathrm{N^{+}}(j)$ is the closed neighborhood of node $j$ that still exists in the remaining network. In order to measure stability, we need to simulate network damage by actually removing nodes from the network and calculating the remaining dominated fraction.

Domination stability depends not only on the fraction of removed nodes, but also on the order in which nodes have been removed from the network. Similarly to many studies in the literature, we consider two damage scenarios: random and targeted node removals. The random node removal strategy models network damage produced by natural causes or errors, while the targeted node removal method reflects the impact of intentional, targeted attacks on a network. In the random damage scenario nodes are removed with equal probability, in random order. In case of targeted attacks, the nodes are removed in degree-ranked order, with highest degrees being removed first. We indicate which strategy we consider in the subscript of stability: $s_{\mathrm{rand}}$ denotes the stability against random damage, and $s_{\mathrm{deg}}$ corresponds to the stability against degree-ranked removal.

\section*{Results}


\subsection*{Stability of Various Fixed Dominating Sets}
We start our analysis by measuring the stability of three different dominating sets, that we use for baseline comparison with our new methods. These are the following:
\begin{itemize}
\item greedy minimum dominating set (MDS) \cite{domination-fundamentals,our-mds,alon-spencer}, where nodes are selected by a sequential greedy search algorithm in order to approximate the actual (NP-hard) smallest dominating set,
\item ``cutoff'' dominating set (CDS) \cite{our-cds}, where all nodes above a degree threshold are selected into set $X$, and the nodes not dominated by any nodes in set $X$ are selected into set $Y$. The dominating set is then given by $X \cup Y$. The degree threshold is selected such that it minimizes the size of the resulting dominating set,
\item degree-ranked dominating set (DDS), where we select all nodes in decreasing order of degree (with random tie-breaking) as dominators until the selected set dominates the entire network.
\end{itemize}
Our first choice is MDS, due to its importance in cost-efficient control of complex networks, and because it provides a high-quality approximation to the actual smallest dominating set. The other methods we have chosen are potentially useful when finding the greedy MDS or solving the binary integer programming equivalent is impractical, e.g., when the adjacency information of the network is incomplete, or the network is too large to run these algorithms in a reasonable amount of time. In these cases heuristic algorithms, such as CDS or DDS can find suboptimal (not the smallest possible), yet small enough dominating sets  that are still useful for practical applications. In particular, the excess nodes selected by these methods may help to increase domination stability.

\begin{figure}[h!]
\centerline{\includegraphics[width=\textwidth]{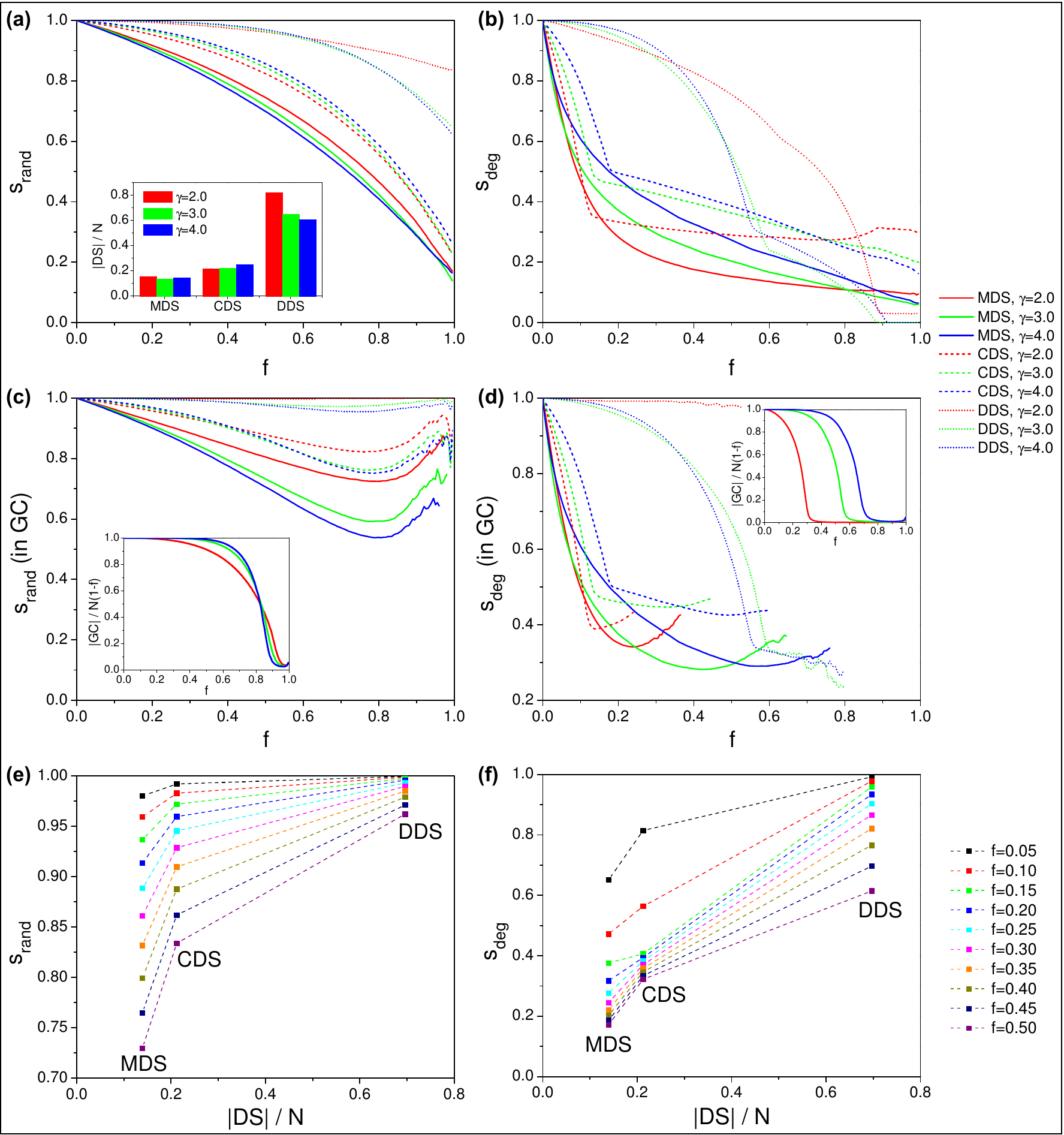}}
\caption{Stability of various dominating sets against random and degree-ranked node removal. Subfigures (a), (c), and (e) show random node removal, (b), (d), and (f) show degree ranked node removal. Subfigures (a) and (b) show stability in the entire network, while (c) and (d) show stability within the remaining giant component. The inset in (a) shows the corresponding sizes of dominating sets, and insets in (c) and (d) show the size of the corresponding giant component. Subfigures (e) and (f) show a correlation between set size and stability, at $\gamma=2.5$. All plots show synthetic scale-free networks, $N=5000$, $\langle k \rangle = 8$, averaged over $200$ network samples.}
\label{fig-1}
\end{figure}

Figure \ref{fig-1} shows the stability against the fraction of removed nodes for MDS, CDS and DDS in the entire remaining network [Fig.~\ref{fig-1}(a), (b)] and in the remaining giant component [Fig.~\ref{fig-1}(c), (d)]. It is clear that the degree-ranked node removal reduces the dominated fraction much faster than the random node removal, because high-degree nodes are more likely to be dominator nodes than low degree nodes. The giant component itself also breaks down much faster, as shown in the insets of Fig.~\ref{fig-1}(c) and (d). However, as long as a giant component exists, it has higher domination stability than the entire network, in both scenarios. The slight increase of stability at high damage rates is a side effect caused by removal of nodes that had lost domination by earlier removals. When the network damage is high, it becomes more likely that these nodes are deleted, causing the dominated fraction of the remaining network to increase. At this point, however, the network is almost completely destroyed and domination stability becomes meaningless.

The stability curves show much more disturbed shapes in degree-ranked removal than random removal, due to the differences in the degree structure of each dominating set. In MDS, there is no preference toward any particular node degree during selection of dominators (besides the natural effect of the greedy selection, where the high-degree nodes provide a larger increase in the number of dominated nodes, hence they are more likely to be selected), which means that removal of high-degree nodes has a smooth (albeit strong) impact on stability. In CDS, we can see a fast initial drop as we remove the very high degree nodes that were specifically selected for dominators (in set $X$), then continuing at a more gentle slope as the dominators from the $Y$ set are removed, since any node that was not dominated by $X$, regardless of degree, may be in set $Y$. On one hand the $Y$ set may seem wasteful in its construction, but with the right degree threshold the size of the CDS is actually very close to the MDS \cite{our-cds}, and the excess nodes provide a fair increase in stability. DDS is the simplest but most inefficient method for finding a dominating set because it selects \textit{all} nodes starting from the highest degrees until all nodes are dominated. However, the resulting redundancy of dominators in the network is providing the highest stability of all three methods.

We can also observe the general tendency that a larger dominating set provides higher stability. At any given fraction of removed nodes, there is a positive correlation between stability and the size of the original dominating set, in both random [Fig.~\ref{fig-1}(a)] and degree-ranked [Fig.~\ref{fig-1}(b)] node removals. We clearly illustrate this correlation in Fig.~\ref{fig-1}(e) and (f), where we show stability as a function of the dominating set size, at various damage levels. This means that the MDS, which is the smallest (most cost-efficient) dominating set, is also the most vulnerable, to both random damage and targeted attacks.

Note, that Fig.~\ref{fig-1} only shows the stability for a certain network type with given degree exponent and uncorrelated networks (where Spearman's $\rho=0$). Stabilities at different values of these parameters are presented in Supplementary Figures S1--S5.

We have also included supplementary videos to illustrate the evolution of domination stability as the network disintegrates, during random node removal (Supplementary Movie 1) and degree-ranked node removal (Supplementary Movie 2).

The main conclusion we can draw is that the extra amount of dominating nodes selected by heuristic methods CDS and DDS, compared to the smaller and more optimal MDS, can effectively increase the stability of domination. However, all three methods are ``fixed'' in the sense that they give only a single possible dominating set size (and corresponding stability) for a given network.

\subsection*{Flexible-Redundancy Dominating Set (frDS)}

In order to overcome the limitations of fixed methods, we must analyze in detail how domination is lost when the network is damaged. First, we realize that loss of domination occurs locally at each node: those nodes that lose all dominators will reduce the domination stability of the network. Therefore, stability can be expressed locally, as the domination \textit{redundancy} of each node. This quantity simply counts how many dominating nodes are within the closed neighborhood of a given node. A large dominating set can successfully increase domination stability, if the extra nodes are distributed in a way that they increase domination redundancy on many nodes. This seems to occur naturally for CDS and DDS, however we cannot guarantee that redundancy was increased in the most optimal way (relative to MDS), nor can we control the number of selected nodes.

We introduce the flexible-redundancy dominating set (frDS) to solve these problems. We explicitly set an average domination redundancy in the network, denoted by $r$, that must be guaranteed by frDS, while aiming for minimum set size. Note, that $r=1$ is equivalent to the minimum dominating set (MDS), and when $r$ is an integer, the frDS is identical to the \textit{h-dominating set} (with $h=r$) studied by Cooper, et al. \cite{hds-cooper}. Finding an frDS is most likely NP-hard, since it is also NP-hard to find an MDS \cite{mds-nphard} or an h-dominating set \cite{hds-nphard}, but we can use a modified greedy algorithm to find an approximation.

The steps of finding an frDS are as follows. First, we assign a domination redundancy requirement, $r(i)$ for each node $i$ as an integer value indicating at least how many dominators node $i$ must have in the dominating set. Given the desired average (non-integer) $r$ value for the entire network, we assign the nearest integer values $\lfloor r \rfloor$ and $\lceil r \rceil$ to each node randomly, such that the network average will be $r$ (the probability of assigning $\lceil r \rceil$ is $r-\lfloor r \rfloor$, which is analogous to a biased coin toss). For the greedy selection we define a dominating potential $p(i)$ as the number of nodes in the closed neighborhood of $i$ that have not yet reached their domination requirement, and therefore selecting node $i$ can help them advance toward their goal. (Note, by definition, the potential of an already selected node is zero.) At each greedy step we select one node with maximum dominating potential (with random tie-breaking), until the requirements of all nodes have been fulfilled. Note, that since dominating potential is an integer number between $0$ and $N$, nodes can be sorted according to their potential in $O(N)$ steps, and it is possible to maintain sortedness after changing the potential of a node in $O(1)$ step (see Supplementary Note 1 for further details and pseudocode). This results in the same computational time complexity as for the greedy MDS approximation, $O(E)$. Also note, that if $r > N$, then the node requirements can never be satisfied, in which case the greedy selection naturally falls back to selecting nodes in degree-ranked order, because at every step every neighbor of a node may be advanced toward its goal.

\subsection*{Flexible-Cost Dominating Set (fcDS)}

When we aim for a desired dominating set size (cost level, i.e., having a limited budget), we can, in principle, aim for the necessary redundancy level in frDS to achieve that desired cost. However, we can further improve stability by considering the expected attack pattern on the network (if the information is available), and optimize the selected dominating set accordingly. For example, if the attack is expected at high-degree nodes, we should avoid selecting many of those nodes as dominators, despite their ability to cover large fractions of the network.

We can optimize our choice of dominators by including the probability of losing each node into the calculation of local stability, which we aim to maximize. First, we assign a strength value $s(i) \in (0,1)$ to each node $i$, which represents the a-priori estimated probability for not losing that node after the attack (i.e., the anticipated attack pattern). Then, we calculate the current domination stability of node $i$ as follows:
\begin{equation}
\mathrm{stability}(\mathrm{DS},i) =
    \begin{cases}
    0 & \mbox{if } \mathrm{DS} \cap N^{+}(i) = \varnothing \\
     1 - \prod_{j \in \mathrm{DS} \cap N^{+}(i)} (1 - s(j)) & \mathrm{otherwise,}
    \end{cases}
\label{eq-stab2}
\end{equation}
which is the probability that node $i$ will remain dominated (not lose all dominators), assuming nodes will be deleted independently; DS denotes the currently selected dominating set. For selecting the next dominator, we choose one that increases the total stability of the network maximally. The total potential increase of stability can be calculated for each node as follows:
\begin{eqnarray}
\mathrm{potential}(i) & = & \sum_{j \in N^{+}(i)} \mathrm{stability}(\mathrm{DS \cup \{i\}},j) - \mathrm{stability}(\mathrm{DS},j) \\
& = & \sum_{j \in N^{+}(i)} (1 - \mathrm{stability}(\mathrm{DS},i)) \cdot s(i).
\label{eq-fcDS-dompot}
\end{eqnarray}
Therefore, we always select a node with maximum potential (with random tie-breaking). Note, that unlike in frDS, the potential here is a non-integer value, thus we can only use comparative sorting to order nodes by potential, which needs $O(N \log N)$ steps. In addition, after selecting each dominator, the stability values have to be recomputed in the selected node's closed neighborhood, and the potentials up to the node's second neighborhood. This involves $O(d^{2})$ nodes, where $d$ is the average degree. Thus, maintaining sortedness of  nodes by their potential requires $O(d^{2} \log N)$ steps after selecting each dominator.

In order to compare stability of fcDS with frDS and other dominating sets, we calculate the ``a-priori'' node strength values as follows: $s(i)=0.5$ for random node removal, and $s(i) = 1 - d(i)/N$ for degree-ranked node removal. Here, we assume the size of the anticipated damage is unknown, thus strength values are expressing relative probabilities only. The strength value for a random damage is arbitrary, as long as it is uniform among the nodes, and it is inversely proportional to node degree in a degree-ranked attack. Further details of fcDS and pseudocode are included in Supplementary Note 2.

\subsection*{Stability Comparison of Dominating Sets}

We seek to answer two main questions in our analysis. First, we want to see how much stability we can achieve by selecting various sizes of dominating sets (in other words, how does the stability scale with larger invested cost of domination). Second, we want to know how much more efficient our methods are compared to the fixed dominating sets, that is, given the same size of dominating set as MDS, CDS, or DDS, how much higher stability can our methods provide.

\begin{figure}[h!]
\includegraphics[width=\textwidth]{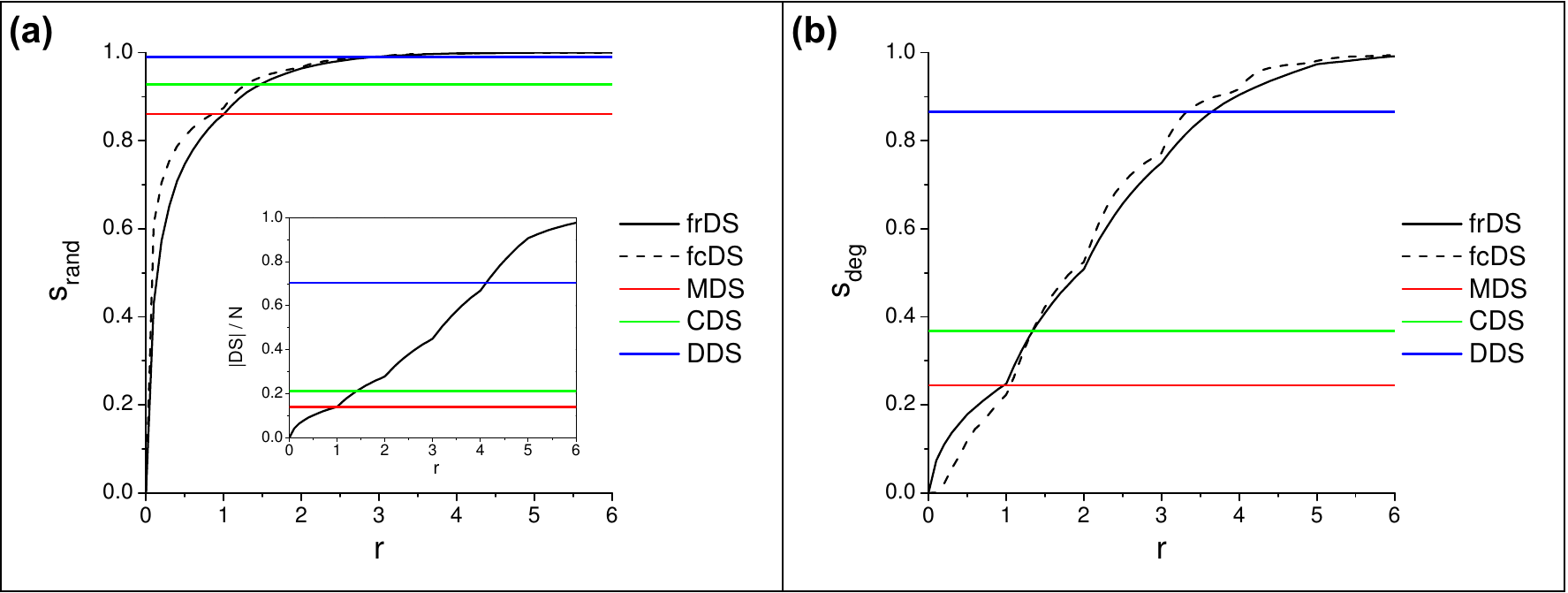}
\caption{Domination stability in frDS and fcDS as a function of domination redundancy. (a) shows random node removal, (b) shows degree-ranked node removal. The inset shows the sizes of the corresponding dominating sets. The size of fcDS is set to match frDS at any given $r$ value. Synthetic scale-free networks, $N=5000$, $\langle k \rangle = 8$, $\gamma=2.5$, $f=0.3$, averaged over $200$ network samples.}
\label{fig-2}
\end{figure}

\begin{figure}[h!]
\includegraphics[width=\textwidth]{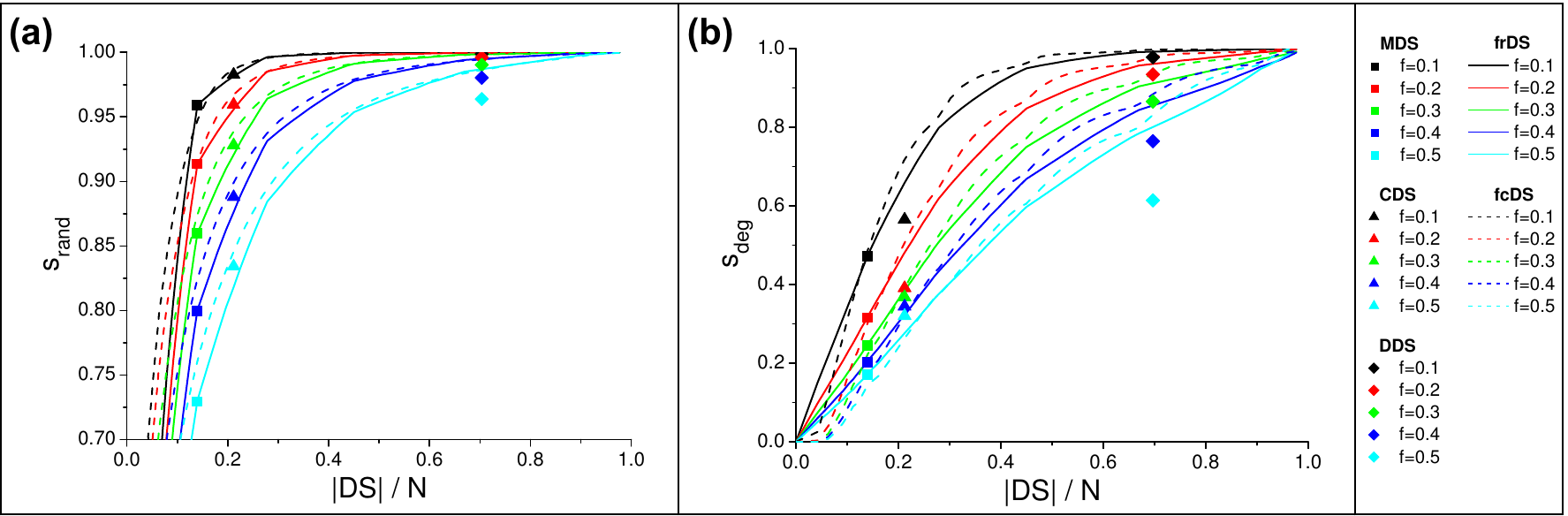}
\caption{Stability of frDS and fcDS as a function of dominating set size (cost) for various network damage fractions. Stabilities of MDS, CDS, and DDS are presented at their corresponding cost values. Subfigure (a) shows random node removal, (b) shows degree-ranked node removal, for synthetic scale-free networks, $N=5000$, $\langle k \rangle = 8$, $\gamma=2.5$, averaged over $200$ network samples.}
\label{fig-3}
\end{figure}

Figures~\ref{fig-2} and \ref{fig-3} show domination stability achieved by frDS and fcDS as a function of redundancy and dominating set size, respectively. Stability achieved by the fixed methods (MDS, CDS, DDS) are also shown at their corresponding cost values for comparison. The general shape of the curves in both figures are similar, since the dominating set size is roughly proportional to redundancy (see Fig.~\ref{fig-2} inset and Supplementary Fig.~S6). In case of random damage, the stability rapidly increases with cost, until the size of MDS is reached, then the curve saturates. There is little advantage in selecting a dominating set larger than approximately twice the size of MDS, because stability is already very close to $1$, even at large damage values. However, in case of degree-ranked damage, there is a steady increase in stability as more nodes are selected as dominators. In both cases, fcDS provides somewhat higher stability than frDS at moderate damage levels, but frDS is more stable at small damage levels. These observations hold across a wide range of network parameters, see Supplementary Figs.~S7 and S8. It is also clear that both frDS and fcDS can provide great flexibility in adjusting the size of the dominating set and stability.

\begin{figure}[h!]
\includegraphics[width=\textwidth]{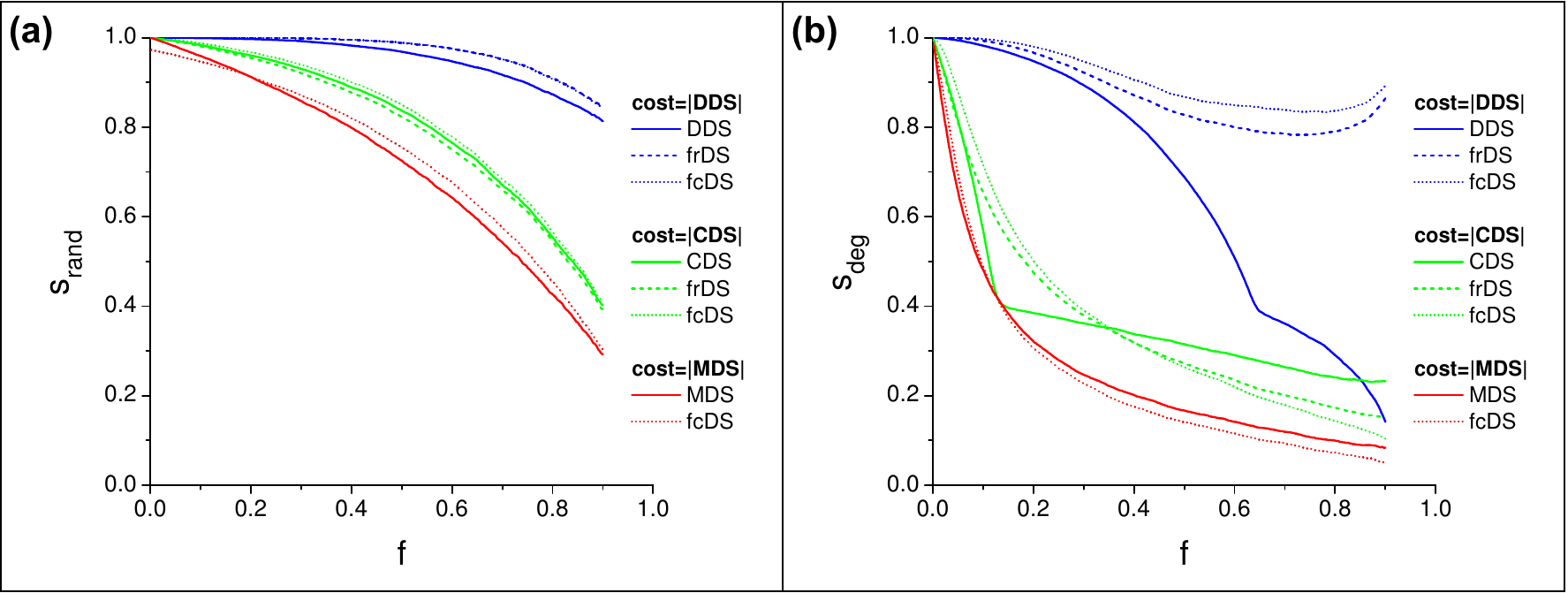}
\caption{Comparison of domination stability at fixed cost levels, as a function of network damage fraction. Stability of frDS and fcDS are plotted at cost values identical to MDS, CDS and DDS. Subfigure (a) shows random node removal, (b) shows degree-ranked node removal, for synthetic scale-free networks, $N=5000$, $\langle k \rangle = 8$, $\gamma=2.5$, averaged over $200$ network samples.}
\label{fig-4}
\end{figure}

The stability of frDS and fcDS at cost levels identical to MDS, CDS, and DDS are presented in Fig.~\ref{fig-4}. Our results show that frDS provides stability values very similar to the fixed methods (in case of MDS, it is identical by definition, thus it is not shown), while fcDS shows a minor improvement in stability. On the other hand, both frDS and fcDS show significant improvement over the fixed methods against degree-ranked attacks, at low network damage fractions. MDS and CDS show a tipping point in damage, where these methods become slightly more effective than frDS or fcDS, but the difference is minimal, and it occurs only at moderate to high network damage ($f \gtrsim 0.3$).

\subsection*{Stability in Real Networks}

We analyze stability of frDS and fcDS, as well as other dominating sets, in several real complex networks, listed in Table~\ref{table-1}. These include an internet peer-to-peer network (p2p-Gnutella08) \cite{stanford}, the power transmission network of continental Europe (ENTSO-E powergrid) \cite{powergrid1,powergrid2}, and one brain graph extracted from MRI data (KKI21-KKI2009-19) \cite{connectome1,connectome2}. Note, that we only use the giant component of these networks. A brief analysis of the degree distribution of Gnutella08 is provided in Supplementary Figs. S15--S17; degree distribution of the powergrid is provided in Supplementary Figs. S18--S21.

The brain graph we analyze here (KKI-21-KKI2009-19) is one of $200$ graphs available from \cite{connectome1}. These graphs have peculiar structural properties, and are very similar to each other. In particular, all brain graphs are very dense: $\langle k \rangle \approx 150$ (Supplementary Fig.~S21); they are all very assortative: $\rho \approx 0.6$ (Supplementary Fig.~S22); and they have very similar degree distributions (see Supplementary Figs.~S23--S25). It is also interesting that the size of MDS is very small, only $3$-$4\%$ network size, while the size of CDS and DDS is very large, around $60\%$ and $100\%$ of network size, respectively (Supplementary Fig.~S26). We attempt to separate the effects of density and assortativity in order to identify their impact on domination stability.

\begin{figure}[h!]
\includegraphics[width=\textwidth]{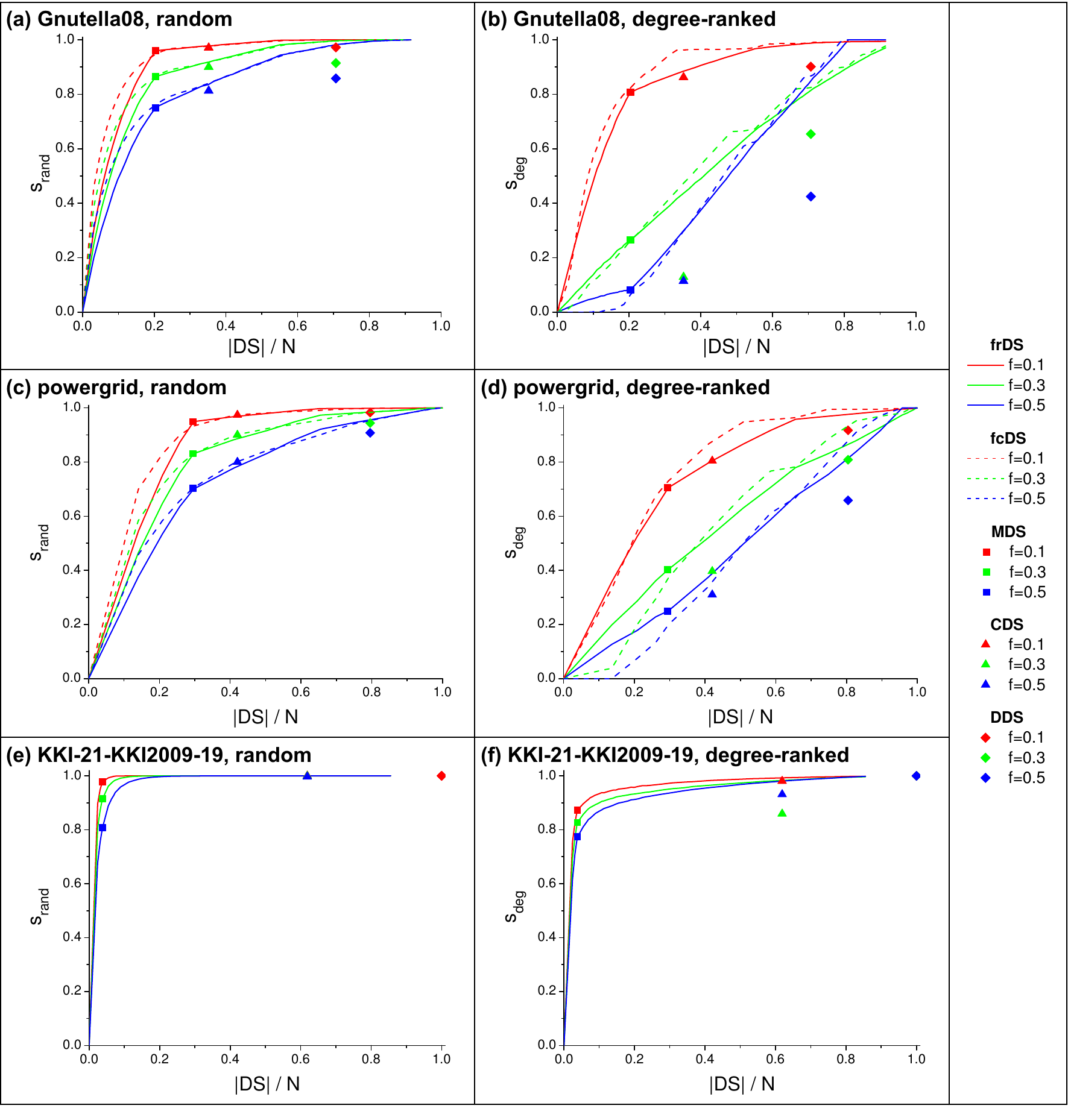}
\caption{Stability of frDS, fcDS and other dominating sets in real networks against random and degree-ranked attacks, for various damage fractions:
(a,b) Gnutella peer-to-peer network; (c,d) ENTSO-E powergrid; (e,f) Brain (MRI) network.
Data is averaged over $20$ independent runs of node removal. See Table~\ref{table-1} for network parameters.}
\label{fig-5}
\end{figure}

Figure~\ref{fig-5} shows domination stability as a function of dominating set size for the real network samples. In general, we see that stability of frDS and fcDS matches the stability of MDS, and exceeds the stability of CDS and DDS, at identical set sizes. In case of Gnutella08 and the powergrid, the stability curves saturate slowly, and the curve shapes are not as smooth as for synthetic scale-free networks, due to having more disturbed (non-scale-free) degree distributions. However, the brain graph shows very high domination stability against both random and targeted attacks. In all cases, the relative advantage of frDS and fcDS over CDS and DDS (i.e., cost-efficiency) remains as high as in synthetic scale-free networks.

We can observe the effects of assortativity separately from other structural properties by artificially changing the network's assortativity, using a biased edge-mixing method (see in \cite{our-cds} and Supplementary Note 3), which rewires the edges in the graph, while keeping the degree sequence unchanged. Using this method we present a brief analysis of dominating set size vs. assortativity in Supplementary Figs.~S27--S29. In general, we see the expected behavior that dominating sets tend to become larger in more assortative networks \cite{our-cds}. Note, that the size of DDS in the brain graph (Supplementary Fig.~S29) being $100\%$ of nodes regardless of assortativity is the result of a particular topological feature; there are a small number of leaves (degree $1$ nodes) connected to degree $2$ nodes, thus DDS has to select al nodes down to degree $2$ (essentially all nodes) to dominate these off-hanging leaves --- a feature left unchanged by edge-mixing.

\begin{figure}[h!]
\includegraphics[width=\textwidth]{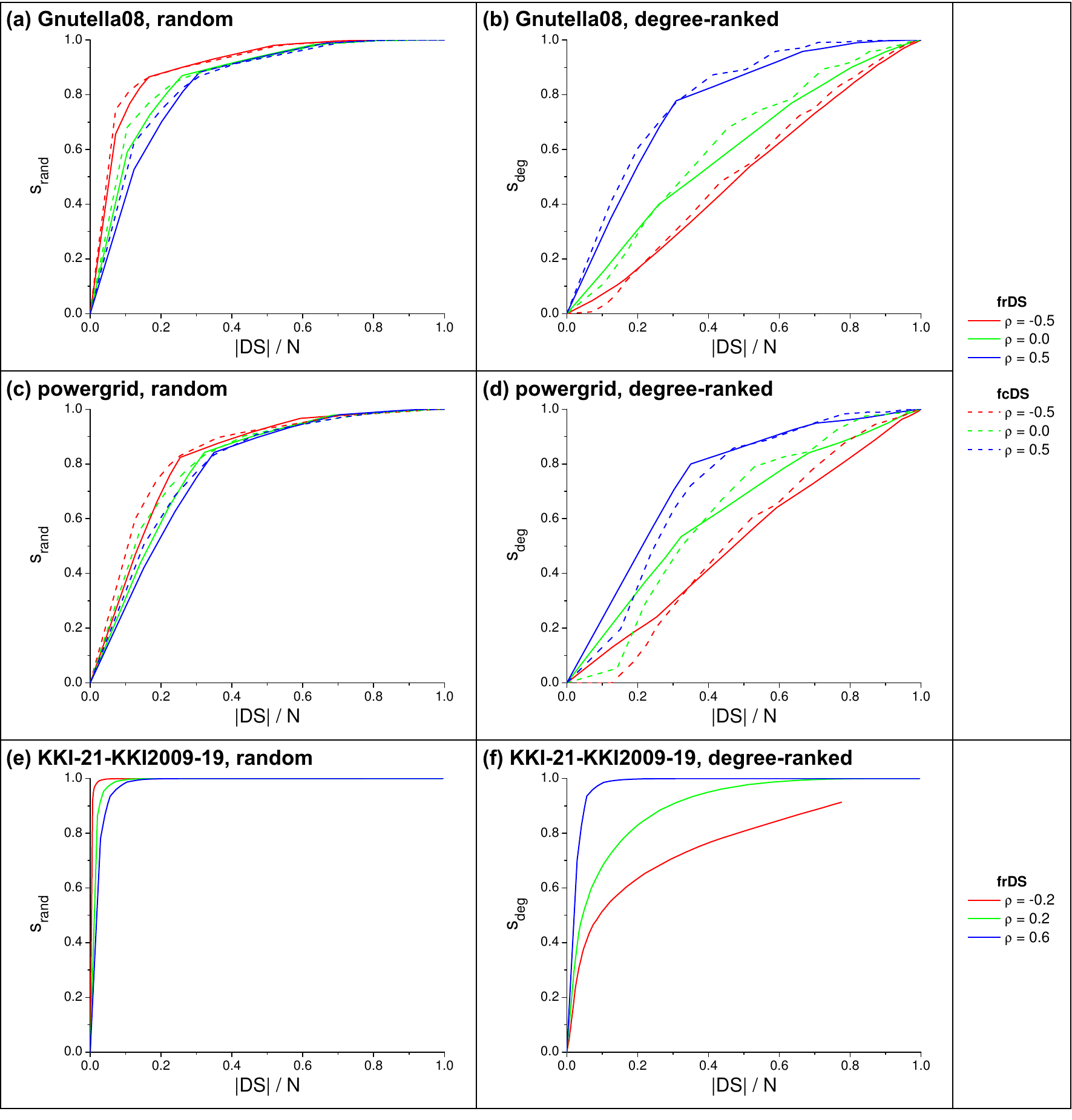}
\caption{Stability of frDS and fcDS in edge-mixed real networks against random and degree-ranked attacks, for various assortativity levels: (a,b) Gnutella peer-to-peer network; (c,d) ENTSO-E powergrid; (e,f) Brain (MRI) network. Network damage fraction $f=0.3$. For (a-d) data is averaged over $50$ independent runs edge mixing and node removal; (e,f) is from a single run. See Table~\ref{table-1} for parameters of the original networks.}
\label{fig-6}
\end{figure}

Figure~\ref{fig-6} presents the effects of assortativity on domination stability. We see an unexpected behavior: as assortativity increases, domination stability decreases against random damage, but increases against an attack on high-degree nodes. We can understand this behavior by considering the effects of assortativity on dominator node degrees. In disassortative networks dominators are mostly high-degree hubs, while in assortative networks dominators have a full range of degrees. Thus, when the network is disassortative and the damage is random, it is less likely to remove high-degree hubs and more likely to remove low degree nodes, the latter rarely being a dominator, leading to increased stability. On the other hand, the result is reversed when high-degree nodes are targeted, in which case we are more likely removing dominators, leading to decreased stability.

Finally, we can conjecture that the outstandingly high domination stability in brain graphs can be attributed to both their high average degree and high assortativity. High average degree results in a highly redundant dominating set (regardless of method) which resists random damage successfully, while high assortativity guarantees that an attack targeted at high degrees leaves the network with plenty of lower-degree dominators.

\subsection*{Partial Flexible-Redundancy Dominating Sets}

There are two possible ways to achieve a certain desired cost (dominating set size) with frDS. Either we aim for the lowest $r$ value that provides the desired cost, or we may choose a larger $r$ value, and use only a fraction of the larger dominating set it provides. In the latter case we would select nodes in the same order as the greedy algorithm picked them. In other words, we can either select a full frDS with small $r$ or a partial frDS with the same size but larger $r$. Figure~\ref{fig-7} shows the comparison of these two cases (see Supplementary Figures S9--S14 for analysis over a wide range of network parameters). The contour curves of fixed stability values are monotonically increasing for larger $r$ values, indicating that the cost for a certain stability level increases if we use partial frDS with higher $r$ values. This also means that using full frDS with the smallest possible $r$ value provides the highest possible stability.

In order to find the needed $r$ value for a desired cost we must look at the relationship between $r$ and the size of the resulting dominating set (see Fig.~\ref{fig-2}(a) inset, and Supplementary Figure S6). The frDS size curve has a complex shape, but it is always monotonically increasing. Therefore, we can use a bisection method for finding the desired $r$ value. Without any assumptions (other than monotonicity) about the size of frDS we must calculate the full frDS for every tested $r$, each taking $O(E)$ time, leading to $O(E \log N)$ time complexity for the entire procedure.

It is also interesting to note that the cost of stability increases slightly for smaller $r$ values when $r<1$, in case of a random damage [in Fig.~\ref{fig-7}(a)]. In this case even the full frDS is providing only a partial dominating set (dominating only a fraction of nodes in the undamaged network). This indicates that $r$ should never be smaller than $1$; if a smaller cost is needed than the one provided by frDS with $r=1$ (which is the MDS by definition), then a partial MDS (given by the greedy MDS algorithm) is a more optimal solution.

\subsection*{Effects of Incorrectly Estimated Damage in fcDS}

For practical applications of fcDS, it is necessary to understand how stability is affected, when the network damage is estimated incorrectly. We can check this effect for a degree-ranked attack by using the following sigmoid strength function for a node with degree $k$:
\begin{equation}
s(k) =\frac{1}{1+e^{\alpha \left(k - \kappa(\alpha, f) \right)}}.
\label{eq-anticipation}
\end{equation}
There are two control parameters for the anticipation. The slope parameter $\alpha \in (-\infty, \infty)$ describes the attack distribution: it expresses whether low degrees ($\alpha<0$) or high degrees ($\alpha>0$) are targeted, and how sharp the difference is between targeted and non-targeted node strengths; parameter $f$ is the anticipated damage fraction. The $\kappa(\alpha, f)$ function gives the threshold for the sigmoid, such that the expected number of lost nodes equals the anticipated damage, $\sum_{k}(1-s(k)) p(k) = f$ (where $p(k)$ is the degree distribution). Note, that $\alpha = \infty$ gives a sharp cutoff selecting all nodes above $\kappa$, corresponding to the actual attack; $0 < \alpha \lesssim 5$ corresponds to an uncertain transition point but correct  anticipation; $a \approx 0$ corresponds to a random guess; $ -5 \lesssim \alpha < 0$ corresponds to an incorrect anticipation (i.e., anticipating attack on low degree nodes, when the attack occurs at high-degree nodes); and $\alpha \ll -5$ is the complete opposite of the actual attack.

\begin{figure}[h!]
\includegraphics[width=\textwidth]{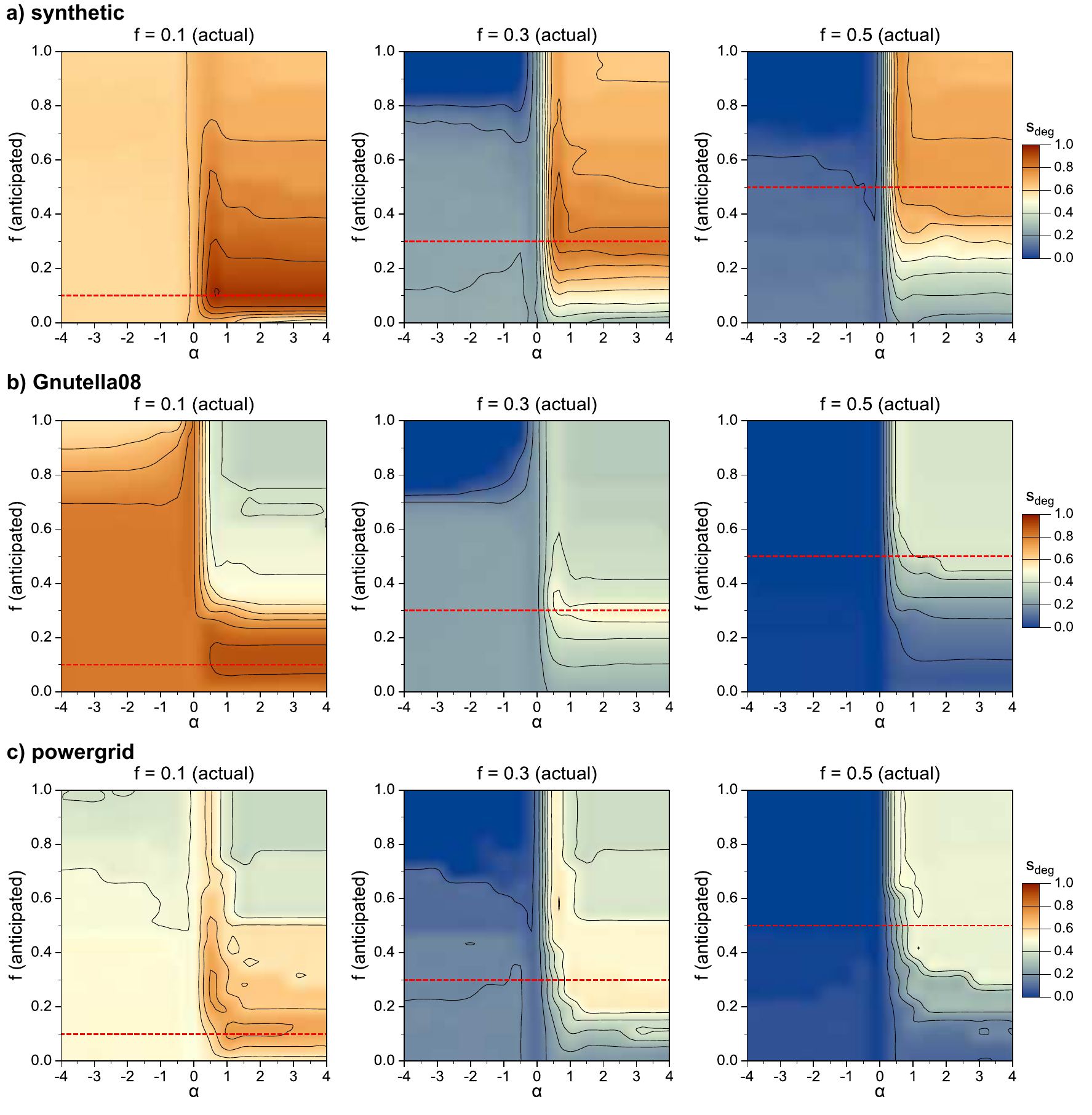}
\caption{Stability of fcDS against degree-ranked node removal as a function of the damage anticipation accuracy: (a) synthetic scale-free network with $N=5000$, $\langle k \rangle = 8$, $\gamma=2.5$; (b) Gnutella peer-to-peer network; (c) ENTSO-E powergrid. The actual damage fraction is indicated above the plots and marked by red dashed lines; the actual degree distribution of the damage corresponds to $\alpha \geq 4$ values.}
\label{fig-8}
\end{figure}

Figure~\ref{fig-8} shows the landscape of stability as a function of the control parameters. As expected, we obtain the highest stability when the attacked degrees and the size of the attack are correctly estimated. For small damage fractions ($f = 0.1$) we lose stability mostly for overestimating the size of the attack, while for moderate ($f = 0.3$) and large ($f = 0.5$) damages we lose stability for incorrectly anticipating which degrees are targeted.

\section*{Discussion}

We must clarify and make a distinction between the prescribed domination redundancy and the actual achieved domination redundancy in a network, when using frDS. The former is the one denoted by the $r$ parameter, while the latter (i.e., the actual number of dominators in the closed neighborhood of a node) can be easily calculated for any given dominating set (not just frDS), and its average always exceeds the prescribed value. For example, even an MDS could have an actual average redundancy of $2.5$ in certain networks, although most nodes would have only one dominator. However, an frDS with $r=2.5$ would guarantee not only that the actual redundancy is at least $2.5$, but also that no nodes will have less than $2$ dominators.

The usage of frDS against degree-ranked or any other targeted attacks seems counter-intuitive, since in frDS, we aim for an overall increased redundancy that is most effective against random damage. However, the greedy algorithm has no preference toward selecting low-degree or high-degree dominators when trying to fulfill domination requirements, and in general, we observe empirically that the selected dominators have a large variability in degrees. This indicates that dominators of a given node may have significantly different degrees, which helps to keep the node dominated even if high degree nodes are targeted by an attack.

In the calculation of node stability in fcDS we assumed that nodes are deleted independently. In a realistic scenario, an attack may have between-node correlations, especially, in spatial graphs (e.g., clustered attack on a power grid). Taking this into count would add more complexity to the calculations, which we postpone for future work. However, it is important to emphasize that even without correlations, the fcDS algorithm can use arbitrary node strength values, irrespective of node degrees, therefore its applicability goes much beyond our studied scenario of a degree-ranked attack.

Currently, the time complexity of fcDS is $O(d^{2} \log N)$ for selecting each dominator node, which makes it prohibitive for very large graphs. In order to speed up the algorithm, the only obstacle we need to overcome is maintaining the sortedness of nodes by their potentials efficiently, which takes $O(\log N)$ steps after each change with comparative sorting. In principle, the potentials could be discretized and assigned to bins (the same optimization we use in frDS), which would lead to $O(E)$ complexity, as long as the bin count remains $O(N)$. However, the effects of such discretization on the dominating set and its stability is unclear, and it would require a thorough analysis to test the method's viability.

We can easily explain that fcDS has a slightly lower stability than frDS at low damage fractions, which we can observe in all graphs, by looking at the effects of incorrect attack anticipation. When the actual damage is very small, we overestimate the damage with our degree-dependent strength formula ($s(i)=1-d(i)/N$), because we assign nonzero probabilities to losing nodes with medium to low degrees. In reality, these nodes will not be deleted in a small targeted attack, thus the overestimated damage causes fcDS to lose stability, dropping slightly below the levels of frDS. This also underlines the need to correctly estimate the size as well as the distribution of the expected attack to achieve the most optimal domination stability.

Finally, we can provide a simple guide for selecting one of our two methods for practical applications. If we have no detailed information about a potential attack, or the network is very large, then frDS is a good choice for providing a dominating set with decent stability against any form of damage (mostly against random damage originating from natural sources), with a short computational time. However, if there is a fixed budget for dominators, or detailed (and reliable) information is available about potential attacks, then fcDS can be used to optimize the selected dominating set for the highest possible stability.

\section*{Methods}

We measure domination stability as an averaged value over an ensemble of networks, using the following procedure. First, a network sample is generated, and its dominating set is calculated by one of the preselected dominating set search algorithms. Then, $m$ nodes are removed from the network, according to a predetermined node removal strategy, where $m/N=f$ is the desired fraction removed from a network with $N$ nodes. Finally, stability is evaluated using Eq.~\ref{eq-def-stab} in the remaining network.

Each node removal strategy is implemented using a sorted list of all nodes in the network; nodes are sorted such that the first $m$ nodes will be removed. For random node removal the list of nodes is shuffled (a random permutation is computed) by the Fisher-Yates algorithm \cite{shuffle}. For degree-ranked node removal the nodes are sorted in decreasing order of degrees (with random tie-breaking).

We generate scale-free network samples using the configuration model \cite{conf-1,conf-2}. First, a discrete power-law degree distribution is constructed for given network size $N$, degree exponent $\gamma$, and average degree $\langle k \rangle$. The degree sequence is then sampled from the degree distribution, and treated as a set of half-links for each node to be connected. Links are realized by randomly (uniformly) selecting any two unconnected half-links, until no more links can be formed. This may result in multiple links between some nodes, but they are treated only as single links, resulting in a small loss of total links. However, the loss is negligible, since we only focus on networks with $\gamma > 2$.

The average degree is controlled by adjusting the minimum degree cutoff $k_{\min}$ of the degree distribution, while the maximum degree cutoff $k_{\max} = \sqrt{N}$. The correct $k_{\min}$ value that yields the desired average degree for the network is obtained from a precomputed lookup table. We have used the same technique in our previous work \cite{our-mds} where we have shown the high level of accuracy achievable with this method. According to our previous notation in \cite{our-mds}, the networks we use here are cCONF networks (abbreviation for configuration model with structural cutoff $k_{\max} = \sqrt{N}$).

\section*{Acknowledgments}
We thank Tao Jia for valuable discussion.
This work was supported in part by grant No. FA9550-12-1-0405 from the U.S.
Air Force Office of Scientific Research (AFOSR) and the Defense
Advanced Research Projects Agency (DARPA), by the Defense Threat
Reduction Agency (DTRA) Award No. HDTRA1-09-1-0049, by the National
Science Foundation (NSF) Grant No. DMR-1246958, by the Army Research
Laboratory (ARL) under Cooperative Agreement Number
W911NF-09-2-0053, by the Army Research Office (ARO) grant
W911NF-12-1-0546, and by the Office of Naval Research (ONR) Grant
No. N00014-09-1-0607. The views and conclusions contained in this
document are those of the authors and should not be interpreted as
representing the official policies either expressed or implied of
the Army Research Laboratory or the U.S. Government.

\section*{Author Contributions}
F.M., N.D., B.K.S. and G.K. designed the research;
F.M. and N.D. implemented and performed numerical experiments and simulations;
F.M., N.D., B.K.S. and G.K. analyzed data and discussed results;
F.M., N.D., B.K.S. and G.K. wrote and reviewed the manuscript.

\section*{Additional Information}
Competing financial interests: The authors declare no competing financial interests.

\newpage

\section*{Figures \& Captions}

\nopagebreak

\begin{figure}[h!]
\includegraphics[width=\textwidth]{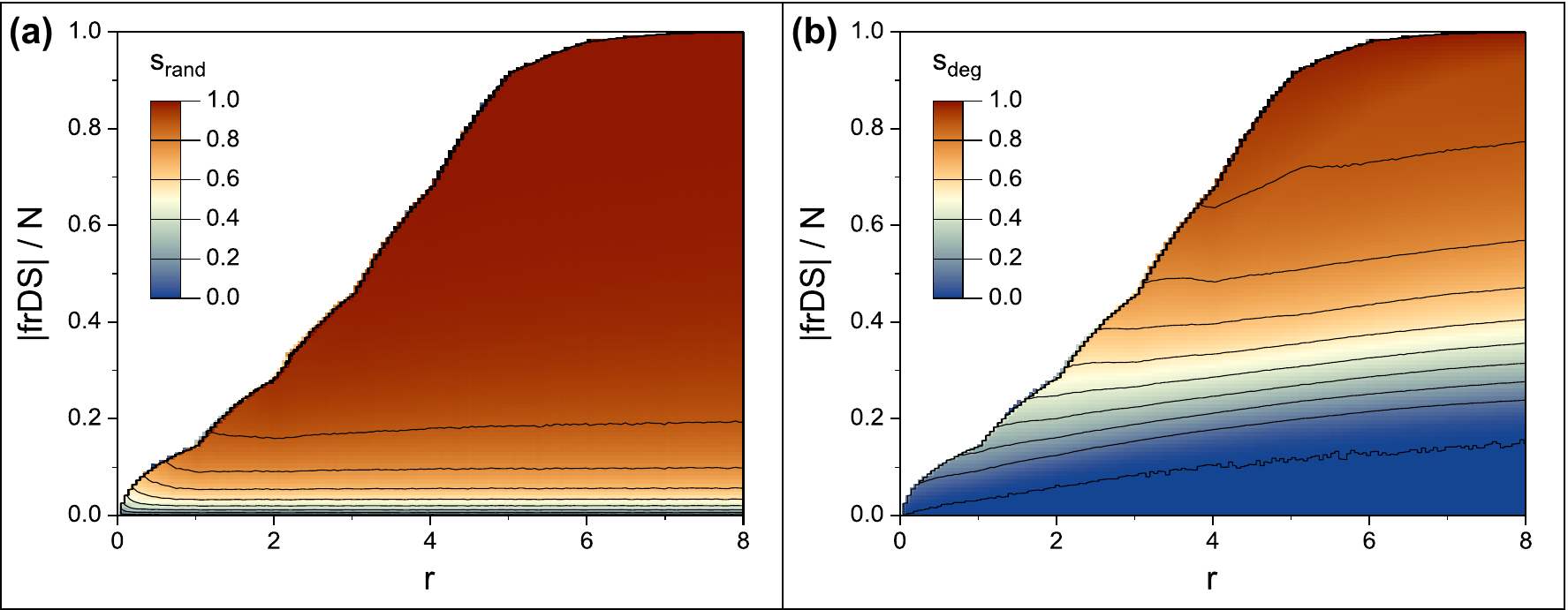}
\caption{Domination stability of partial frDS as a function of domination redundancy and dominating set size. The plotted area is bounded by the size of the full frDS at any given $r$. Subfigure (a) shows random node removal, (b) shows degree-ranked node removal, for synthetic scale-free networks, $N=5000$, $\langle k \rangle = 8$, $\gamma=2.5$, $f=0.3$, averaged over $50$ network samples.}
\label{fig-7}
\end{figure}

\clearpage

\begin{table}
\begin{center}
\begin{tabular}{| c | c | c | c | c | c | c | c |}
\hline
Name & Source & N & $k_{\min}$ & $k_{\max}$ & $\langle k \rangle$ & Spearman's $\rho$ \\ \hline
Gnutella08 & \cite{stanford} & $6299$ & $1$ & $97$ & $6.60$ & $0.03$ \\ \hline
powergrid & \cite{powergrid1,powergrid2} & $1494$ & $1$ & $13$ & $2.89$ & $-0.18$ \\ \hline
KKI-21-KKI2009-19 & \cite{connectome1,connectome2} & $712098$ & $1$ & $6505$ & $138.2$ & $0.62$ \\
\hline
\end{tabular}
\end{center}
\caption{Parameters of real networks used in our analysis. The data refers exclusively to the giant component.}
\label{table-1}
\end{table}

\clearpage

\renewcommand{\thefigure}{S\arabic{figure}}\setcounter{figure}{0}
\renewcommand{\theequation}{S\arabic{equation}}\setcounter{equation}{0} 
\renewcommand{\thesection}{S\arabic{section}}\setcounter{section}{0}
\renewcommand{\thesubsection}{\thesection.\arabic{subsection}}

\setcounter{page}{1}

\newgeometry{left=1.0in, right=1.0in}
\newgeometry{top=0.5in, bottom=0.5in}

\begin{center}
\LARGE 
Building Damage-Resilient Dominating Sets in Complex Networks against Random and Targeted Attacks \\ \vspace{0.5cm}
{\bf  Supplementary Information} \\ \vspace{0.5cm}
\large
F. Moln\'{a}r Jr.$^{1,2,}\footnote{E-mail: molnaf@rpi.edu}$, N. Derzsy$^{1,2}$, B. K. Szymanski$^{2,3}$, G. Korniss$^{1,2}$
\end{center}

\begin{flushleft}
$^{\bf{1}}$ Department of Physics, Rensselaer Polytechnic Institute,
110 8$^{th}$ Street, Troy, NY, 12180-3590 USA \\
$^{\bf{2}}$ Social Cognitive Networks Academic Research Center,
Rensselaer Polytechnic Institute, 110 8$^{th}$ Street, Troy, NY, 12180-3590 USA \\
$^{\bf{3}}$ Department of Computer Science,
Rensselaer Polytechnic Institute, 110 8$^{th}$ Street, Troy, NY, 12180-3590 USA \\
\end{flushleft}

\setcounter{figure}{0}

\renewcommand{\figurename}{\bf Supplementary Figure}
\renewcommand{\tablename}{\bf Supplementary Table}
\renewcommand\refname{Supplementary References}

\makeatletter 
\renewcommand{\thefigure}{S\@arabic\c@figure}
\makeatother

\makeatletter
\renewcommand{\thetable}{S\@arabic\c@table}
\makeatother

\makeatletter
\renewcommand{\thealgorithm}{S\@arabic\c@algorithm}
\makeatother

\makeatletter
\setlength{\@fptop}{0pt}
\setlength{\@fpbot}{0pt plus 1fil}
\makeatletter

\section*{Supplementary Note 1\\
\large Algorithm for Finding a Flexible-Redundancy Dominating Set (frDS)}

The algorithm for finding a flexible-redundancy dominating set (frDS) is based on greedy search. At each step we add one node to the dominating set, that helps the maximum number of nodes to advance toward their required domination goals. There are several variables that we must define and track for each node.

First, we define the domination requirement $r(i)$ as the number of required dominators for node $i$ among its closed neighbors. This value is calculated and assigned randomly for each node before the search begins. The requirement is either $\lfloor r \rfloor$ or $\lceil r \rceil$ (where $r$ is the global requirement for the entire network), the probability for the latter is exactly the fractional part of $r$ (that is, $r-\lfloor r \rfloor$). Note, that $r(i)$ can be zero if $r<1$, and it is also possible that $r > d(i)+1$ (where $d(i)$ is the degree of node $i$), in which case all nodes in the closed neighborhood are required to be in the dominating set.

Second, we define $score(i)$ as the current number of dominators of node $i$ at any given step. Initially, $score(i)=0$ for every node, and it increases by one in the closed neighborhood of the selected node.

Finally, we track the dominating $potential(i)$ of node $i$, which counts how many nodes in the closed neighborhood of $i$ have not yet reached their domination requirement. Specifically, $potential(i) = \sum _{j \in N^{+}(i)} I[score(j) < r(j)]$, where $N^{+}(i)$ is the closed neighborhood of $i$ and $I[x]$ is an indicator function that returns $1$ if $x$ is true and $0$ else. In other words, the potential is the number of nodes in the closed neighborhood that can be advanced toward their goal by selecting $i$ as the next dominator. The greedy search is based on this quantity: at every step we select a node with maximum potential (with random tie-breaking among the candidates).

The key to implementing the algorithm with optimal time complexity is the use of an efficient data structure for maintaining a list of nodes sorted by their potentials. Note that the potential is an integer value between $0$ and $N+1$, therefore we can use bucket-sort for initial sorting. We assign one bucket for each possible potential value, and we implement each bucket by a hashed set. This way we can add or remove a node from any bucket in $O(1)$ step, therefore we can perform the initial sorting in $O(N)$ and maintain sortedness in $O(1)$ step after any single change in a node's potential. 

The time complexity of the algorithm can be found by analyzing the changes in scores and potentials of nodes. The initial calculation of potentials requires a loop over all nodes' all neighbors. Assuming we can enumerate the neighbors of node $i$ in $d(i)$ steps, this calculation takes $\sum _{i \in V(G)} 1+d(i) = 2E + N = O(E)$ steps. Then in the main loop one node is selected at every step, which increases the score of the selected node and its neighbors by one. In principle, the scores could increase until all nodes are selected (e.g., when $r > N$), therefore again all nodes' all neighbors are processed, taking $O(E)$ steps. However, during this procedure, there are additional steps for updating the node potentials.  Some (usually all) nodes will reach their predefined requirement at one point or another, after which the dominating potentials change. We count these changes as follows. Initially, all nodes can increase all their neighbors' score toward their requirement (including the nodes themselves), therefore the initial sum of potentials is $\sum _{i \in V(G)} 1+d(i) = 2E + N$, or less, if some nodes have zero requirement. The potential of a node can either be reduced by one if a neighbor reaches its requirement (and thus that neighbor can no longer be advanced to its goal by the current node), or it becomes zero by definition if the node is actually selected. At most, there are $2E+N=O(E)$ changes (reductions) of potentials, each computed in $O(1)$ time (maintaining sortedness of nodes after each change), therefore during the procedure there are at most $O(E)$ additional steps for updating node potentials. This means the entire algorithm runs in $O(E)$ steps. Note that in sparse networks, $O(E) = O(N)$.

\begin{algorithm}
\caption{Find an frDS}
\label{alg-1}
\begin{algorithmic}
\Procedure {frDS}{$G$: graph, $r$: domination redundancy}
 \State $finished \gets 0$

 \ForAll {$i \in V(G)$} \Comment{initialization of $score$, $r$, and $potential$}
  \State $score(i) \gets 0$
  \State $potential(i) \gets 0$
  \If {$Random(0,1) < r - \lfloor r \rfloor$}
   \State $r(i) \gets \lceil r \rceil$
  \Else
   \State $r(i) \gets \lfloor r \rfloor$
  \EndIf
  \If {$score(i) \geq r(i)$}
   \State $finished \gets finished + 1$
  \EndIf
 \EndFor

 \ForAll {$i \in V(G)$} \Comment{initial calculation of potentials}
  \ForAll {$j: (i,j) \in E(G)$}
   \If {$score(j) < r(j)$}
    \State $potential(i) \gets potential(i) + 1$
   \EndIf
  \EndFor
  \If {$score(i) < goal(i)$}
   \State $potential(i) \gets potential(i) + 1$
  \EndIf
 \EndFor

 \While{$finished < |V(G)| \land max(potential) > 0$} \Comment{main loop}
  \State $k:=$ random node with maximum potential \Comment{greedy step}
  \State Add $k$ to Dominating Set \Comment{construct the output}
  \State $score(k) \gets score(k) + 1$ \Comment{count self-domination}
  \State $potential(k) \gets 0$ \Comment{remove $k$ from further consideration}
  \State $change \gets score(k) = r(k)$ \Comment{requirement of $k$ reached in this iteration?}
  \If {$change$}
   \State $finished \gets finished + 1$
  \EndIf

  \ForAll {$j: (j,k) \in E(G)$} \Comment{update neighbors of $k$}
   \If {$change$} \Comment{neighbors cannot increase $score(k)$ any more}
    \State $potential(j) \gets max(0, potential(j) - 1)$
   \EndIf
   \State $score(j) \gets score(j) + 1$ \Comment{$k$ adds domination score to all its neighbors}
   \If {$score(j) = r(j)$} \Comment{requirement reached for the neighbor?}
    \State $finished \gets finished + 1$
    \State $potential(j) \gets max(0, potential(j) - 1)$
    \ForAll {$x: (x,j) \in E(G)$} \Comment{update potentials of second neighbors}
     \If {$x \neq k$} \Comment{skip when second neighbor is $k$}
      \State $potential(x) \gets max(0, potential(x) - 1)$
     \EndIf
    \EndFor 
   \EndIf
  \EndFor

 \EndWhile
\EndProcedure
\end{algorithmic}
\end{algorithm}

\clearpage

\section*{Supplementary Note 2\\
\large Algorithm for Finding a Flexible-Cost Dominating Set (fcDS)}

The fcDS algorithm is also a form of greedy search, since it builds the dominating set by selecting one node at a time with maximum potential, similarly to the frDS algorithm. However, in this method the potential is calculated from the changes in probability of losing all dominators for the nodes in the neighborhood of the given node.

First, we define $strength(i)$ for each node $i$ as an input ($0<strength(i)<1$), which defines the probability of not losing node $i$ after the anticipated damage:
\begin{equation}
strength(i) := \Pr (i \mathrm{\ is\ not\ lost}). 
\label{eq-strength}
\end{equation}
We also keep a record of $instability(i)$ for each node $i$, which is defined as the probability of losing all dominators after the damage has occured: 
\begin{equation}
instability(i) = \prod_{j \in \mathrm{DS} \cap N^{+}(i)} 1 - strength(j).
\label{eq-instability}
\end{equation}
Initially, $instability(i)=1.0$ for all $i$. The $potential(i)$ of node $i$, which is used in the greedy node selection, is calculated as the sum of the changes in instabilities over the closed neighborhood of node $i$, if $i$ was selected:
\begin{equation}
\begin{split}
potential(i) &= -\sum_{j \in N^{+}(i)} instability(j) \Pr(i \mathrm{\ is \ lost}) - instability(j) \\ 
&= -\sum_{j \in N^{+}(i)} instability(j) \left[ \Pr(i \mathrm{\ is \ lost}) - 1 \right] \\
&= \sum_{j \in N^{+}(i)} instability(j)  strength(i).
\end{split}
\label{eq-potential}
\end{equation}
Note, that the negative sign is manually inserted to make the potential a positive value, for practical reasons. Without it, the change in instabilities would be negative, because by each node selection the stability always increases. 

With the definition above, we select a node with maximum potential at each greedy step. After the node has been selected and added to the dominating set, the instabilities in the closed neighborhood, and the potentials for all nodes in the second neighborhood of the selected node must be recalculated, and the nodes must be sorted again based on the new potentials. Since the potentials are non-integer values (and cannot be mapped to integers) we can only use comparative sorting, where it takes $O(\log N)$ steps to find the new place for each node in the list. With a simple approximation for sparse networks, a node in a network with average degree $d$ will have $O(d^{2})$ nodes in its second neighborhood, therefore the selection of each dominator involves $O(d^{2}\log N)$ steps.

\begin{algorithm}
\caption{Find an fcDS}
\label{alg-1}
\begin{algorithmic}
\Procedure {fcDS}{$G$: graph, $strength$: array, $c$: number of nodes to select}

 \ForAll {$i \in V(G)$} \Comment{initialization of $instability$ and $potential$}
  \State $instability(i) \gets 1.0$
  \State $potential(i) \gets (degree(i)+1)(1-strength(i))$
 \EndFor

 \For {$a \gets 1\ ...\ c$} \Comment{$a$ simply counts the output}
  \State $k \gets$ random node with maximum potential \Comment{greedy selection}
  \State Add $k$ to Dominating Set \Comment{construct the output}

  \State $S \gets \varnothing$ \Comment{set of nodes whose potential must be updated}

  \State $instability(k) \gets instability(k) (1 - strength(k))$ \Comment {update self instability}
  \ForAll {$j: (k,j) \in E(G)$}
   \State $instability(j) \gets instability(j) (1 - strength(k)$ \Comment {update instability of neighbors}
   \State $S \gets S \cup \{j\}$ \Comment {request potential update for $j$}
   \ForAll {$i: (j,i) \in E(G)$}
    \State $S \gets S \cup \{i\}$ \Comment {request potential update for second neighbors}
   \EndFor
  \EndFor

  \ForAll {$i \in S$} \Comment{update potentials}
   \State $potential(i) \gets 0$
   \If {$i \notin$ Dominating Set}
    \ForAll {$j: (i,j) \in E(G)$} 
     \State $potential(i) \gets potential(i) + instability(j) strength(i)$
    \EndFor
    \State $potential(i) \gets potential(i) + instability(i) strength(i)$
   \EndIf
  \EndFor

 \EndFor
\EndProcedure
\end{algorithmic}
\end{algorithm}

\clearpage

\section*{Supplementary Note 3\\
\large Measuring and Controlling Assortativity}

Assortativity of a network, defined loosely, is the tendency that nodes of similar degrees are connected to each other. We measure assortativity of a network using Spearman's $\rho$ \cite{s_Spearman_1904}, which has been shown recently \cite{s_Litvak_2013} to be a much more accurate measure of assortativity than Newman's assortativity coefficient \cite{s_Newman_2003,s_Newman_2002}. Spearman's $\rho$ values have a range of $(-1, 1)$, ranging from completely disassortative to completely assortative. Note, that a network with $\rho=0$ is called an uncorrelated network.

In our recent work \cite{s_our-cds} we have proposed a method to control the assortativity of a network using a Markov-chain of double-edge swaps \cite{s_Viger_2005}, guided by biased edge-swap acceptance probabilities. These swaps preserve the degree sequence of a network, but change its assortativity. We provide here an overview of this method again for the reader's convenience.

In our method, we look for randomly (uniformly) selected pairs of edges that have four distinct nodes and no common incident edges. These edge pairs allow for a double-edge swap. However, instead of accepting all swaps, we only accept them with the following probability:
\begin{equation}
\Pr(\textrm{accept})=
    \begin{cases}
    p & \mbox{if } p>0 \mbox{ and the swap makes the network more assortative} \\
    -p & \mbox{if } p<0 \mbox{ and the swap makes the network more dissortative} \\
    1-|p| & \mathrm{else,}
    \end{cases}
\label{eq-pr-accept}
\end{equation}
where $-1 < p < 1$ is a parameter that controls the acceptance ratio of assortative and dissortative swaps. A swap is classified as assortative or dissortative if it increases or decreases the assortativity coefficient \cite{s_Newman_2003,s_Newman_2002} of the network, which can be evaluated very quickly. 

Using our guided edge-mixing, we can reach a wide range of $\rho$ values for any given network; however, determining the correct $p$ value for a desired $\rho$ is nontrivial. Due to the random nature of the mixing procedure, the resulting value of $\rho$ is a random variable. The mean of $\rho$ monotonically increases as $p$ increases, allowing us to use a randomized bisection search to find the needed $p$ for a desired $\rho$. 

\clearpage


\begin{figure}[h!]
\includegraphics[width=\textwidth]{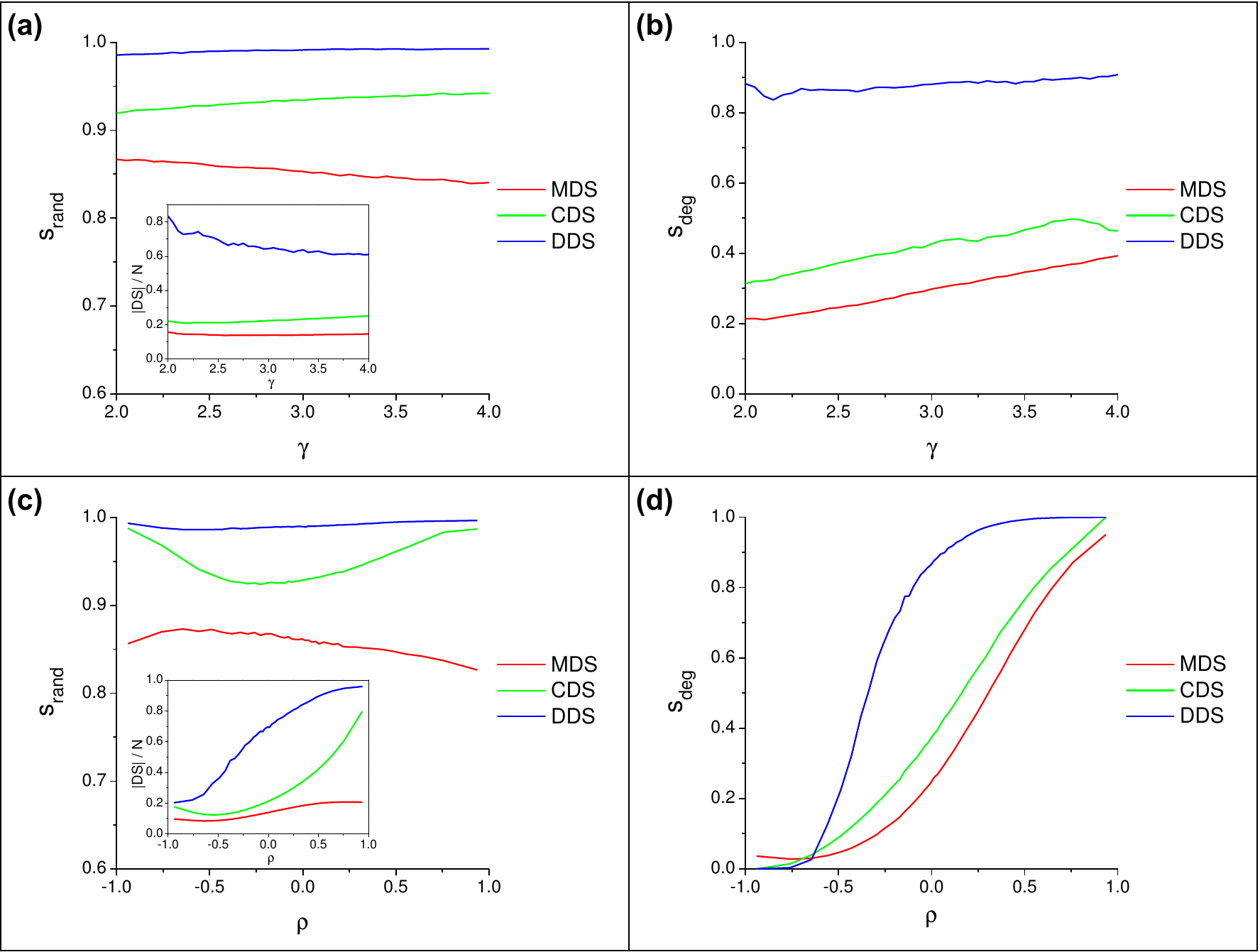}
\caption{Stability of dominating sets vs. power-law degree exponent $\gamma$ and Spearman's $\rho$ assortativity measure. (a) and (c) present random node removal, (b) and (d) show degree-ranked node removal. The insets illustrate the sizes of the corresponding dominating sets. In (a) and (b): $\rho=0.0$; in (c) and (d): $\gamma=2.5$. Common parameters: $N=5000$, $\langle k \rangle = 8$, $f = 0.3$. Results are averaged over $200$ network samples.}
\label{fig-s1}
\end{figure}

\begin{figure}[h!]
\begin{center}
\includegraphics[width=\textwidth]{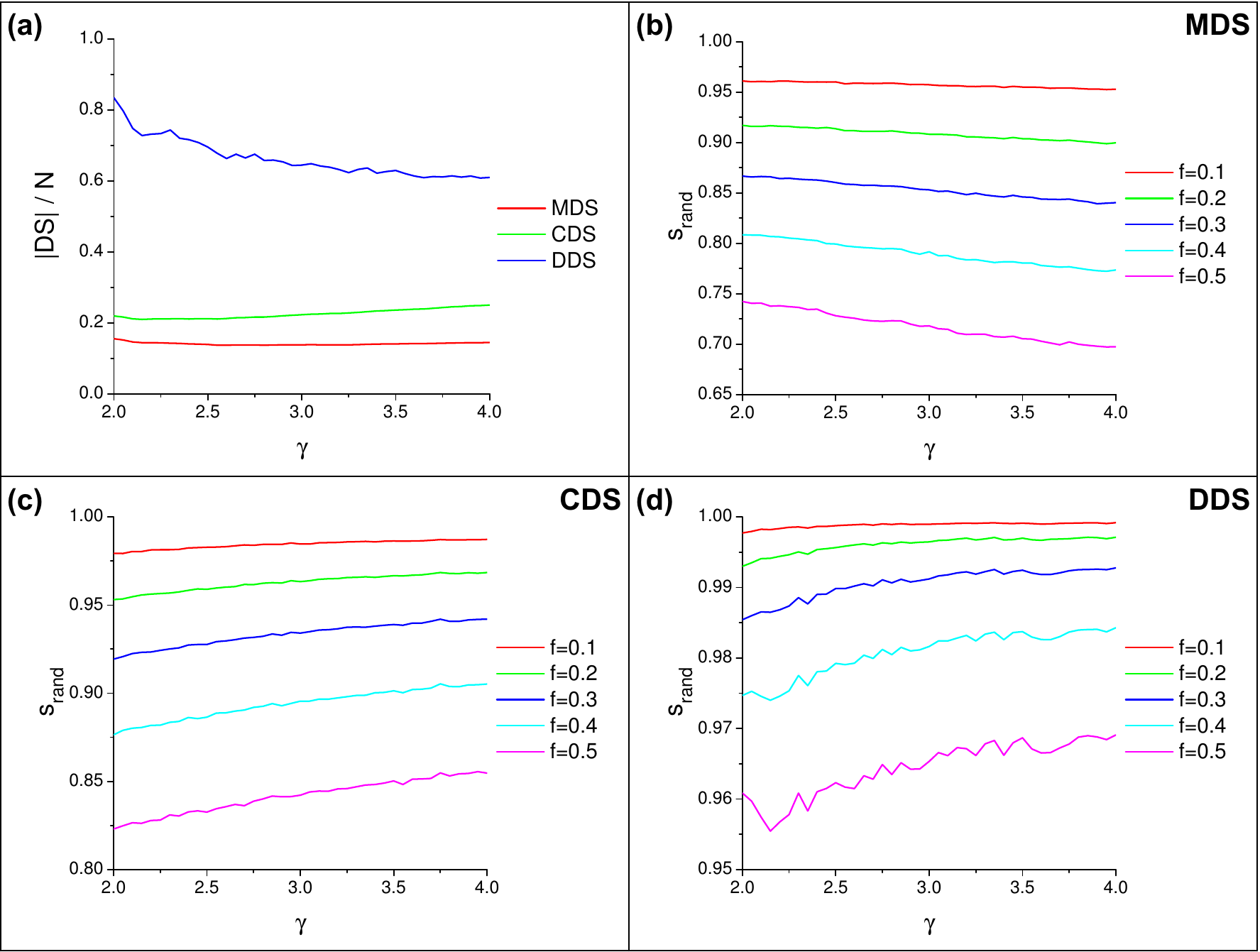}
\caption{Comparison of size and stability of dominating sets vs. power-law degree exponent, at various fractions of random node removal. Synthetic networks, $N=5000$, $\langle k \rangle = 8$, $\rho=0.0$. Results are averaged over $200$ network samples.}
\label{fig-s2}
\end{center}
\end{figure}

\begin{figure}[h!]
\begin{center}
\includegraphics[width=\textwidth]{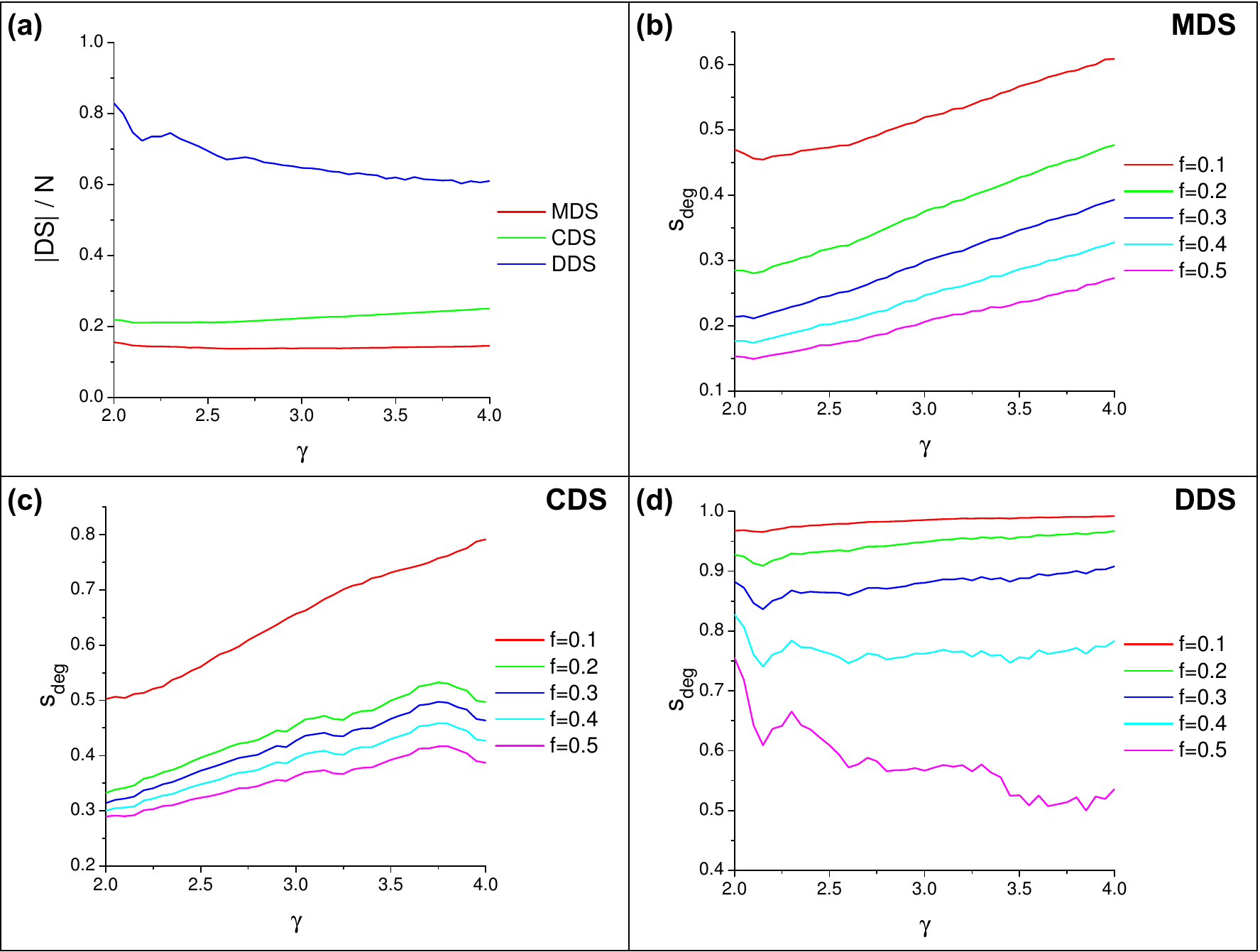}
\caption{Comparison of size and stability of dominating sets vs. power-law degree exponent at various fractions of degree ranked node removal. Synthetic networks, $N=5000$, $\langle k \rangle = 8$, $\rho=0.0$. Results are averaged over $200$ network samples.}
\label{fig-s3}
\end{center}
\end{figure}

\begin{figure}[h!]
\begin{center}
\includegraphics[width=\textwidth]{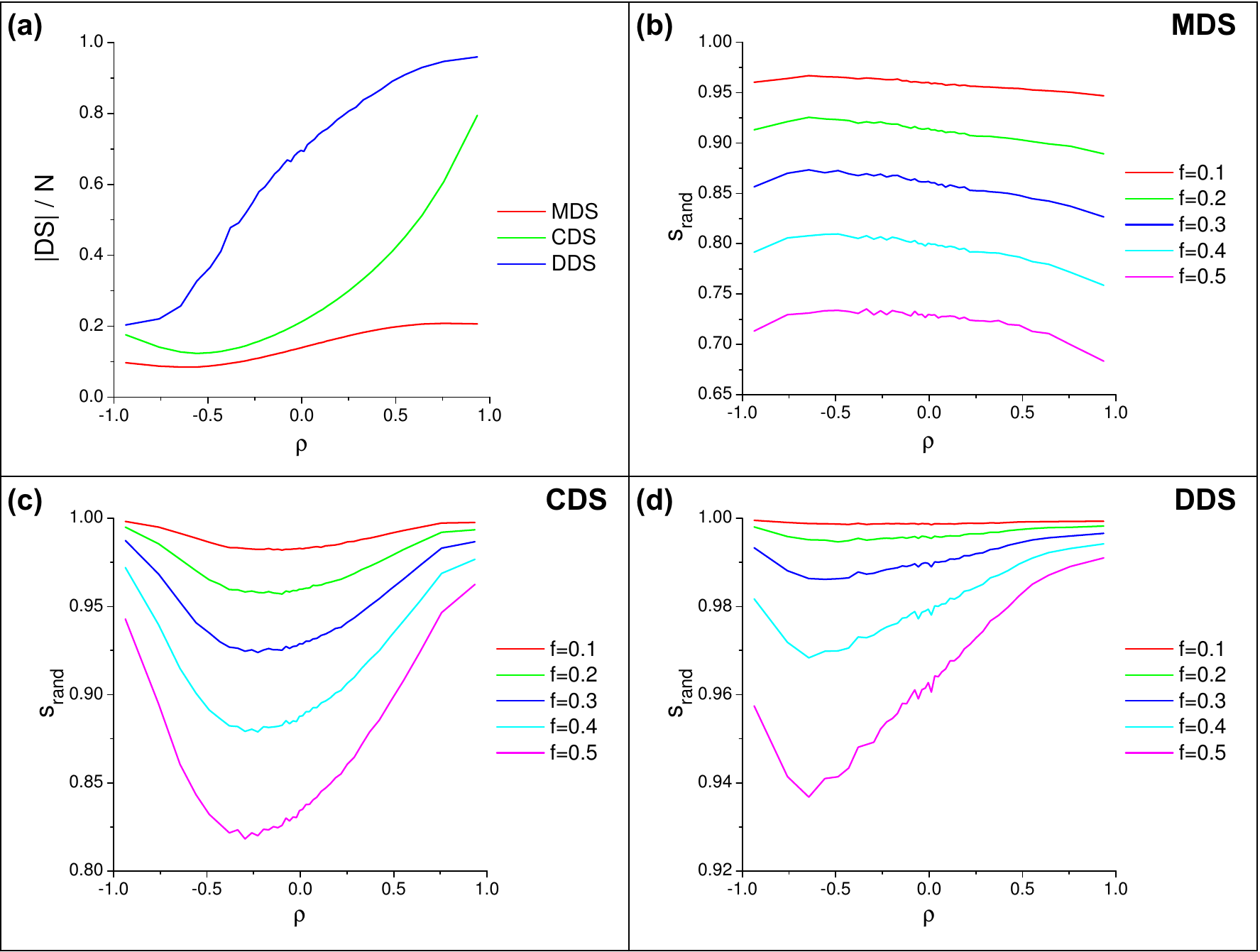}
\caption{Comparison of size and stability dominating sets vs. assortativity at various fractions of random node removal. Synthetic networks, $N=5000$, $\langle k \rangle = 8$, $\gamma=2.5$. Results are averaged over $200$ network samples.}
\label{fig-s4}
\end{center}
\end{figure}

\begin{figure}[h!]
\begin{center}
\includegraphics[width=\textwidth]{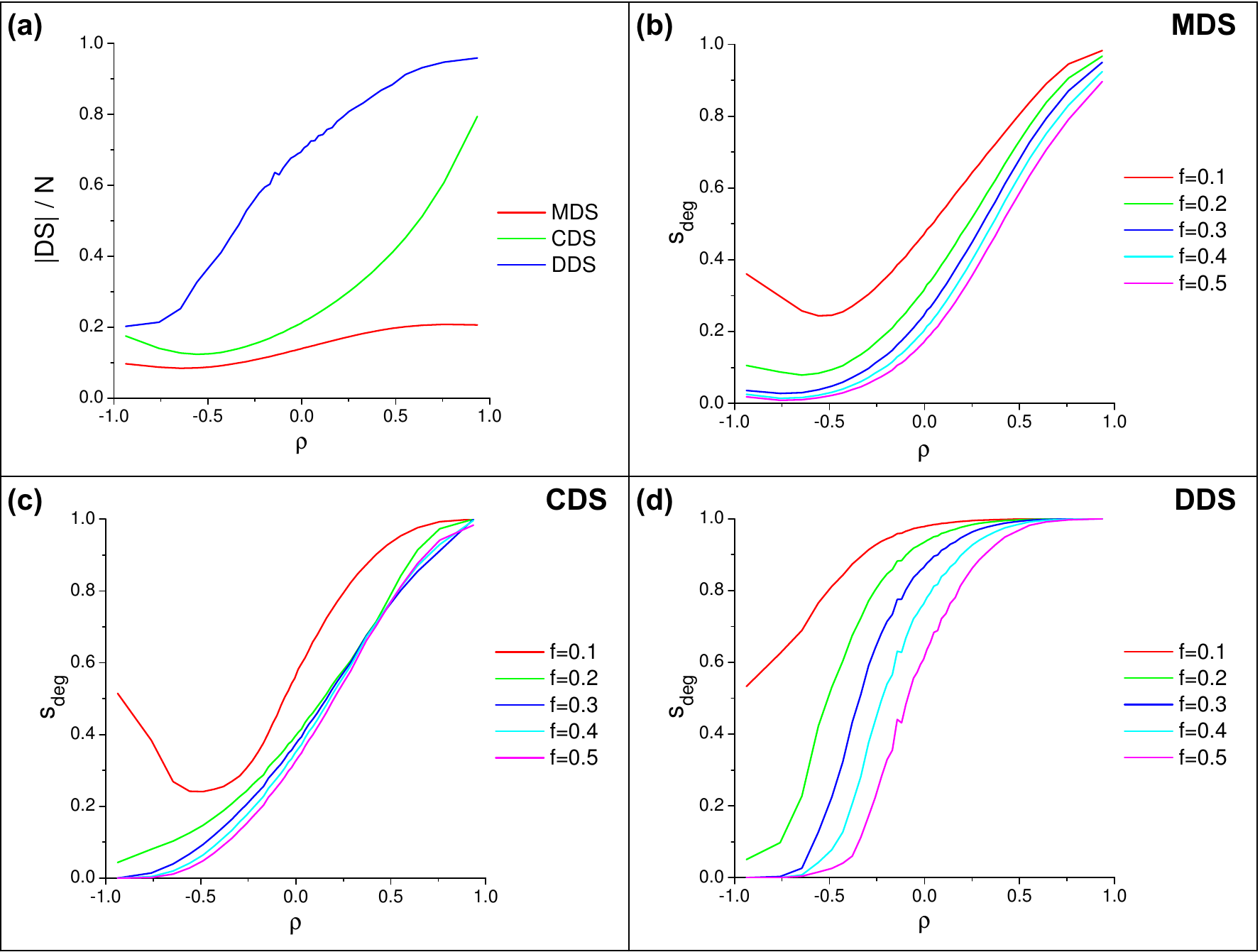}
\caption{Comparison of size and stability dominating sets vs. assortativity at various fractions of degree-ranked node removal. Synthetic networks, $N=5000$, $\langle k \rangle = 8$, $\gamma=2.5$. Results are averaged over $200$ network samples.}
\label{fig-s5}
\end{center}
\end{figure}

\clearpage

\begin{figure}[h!]
\includegraphics[width=\textwidth]{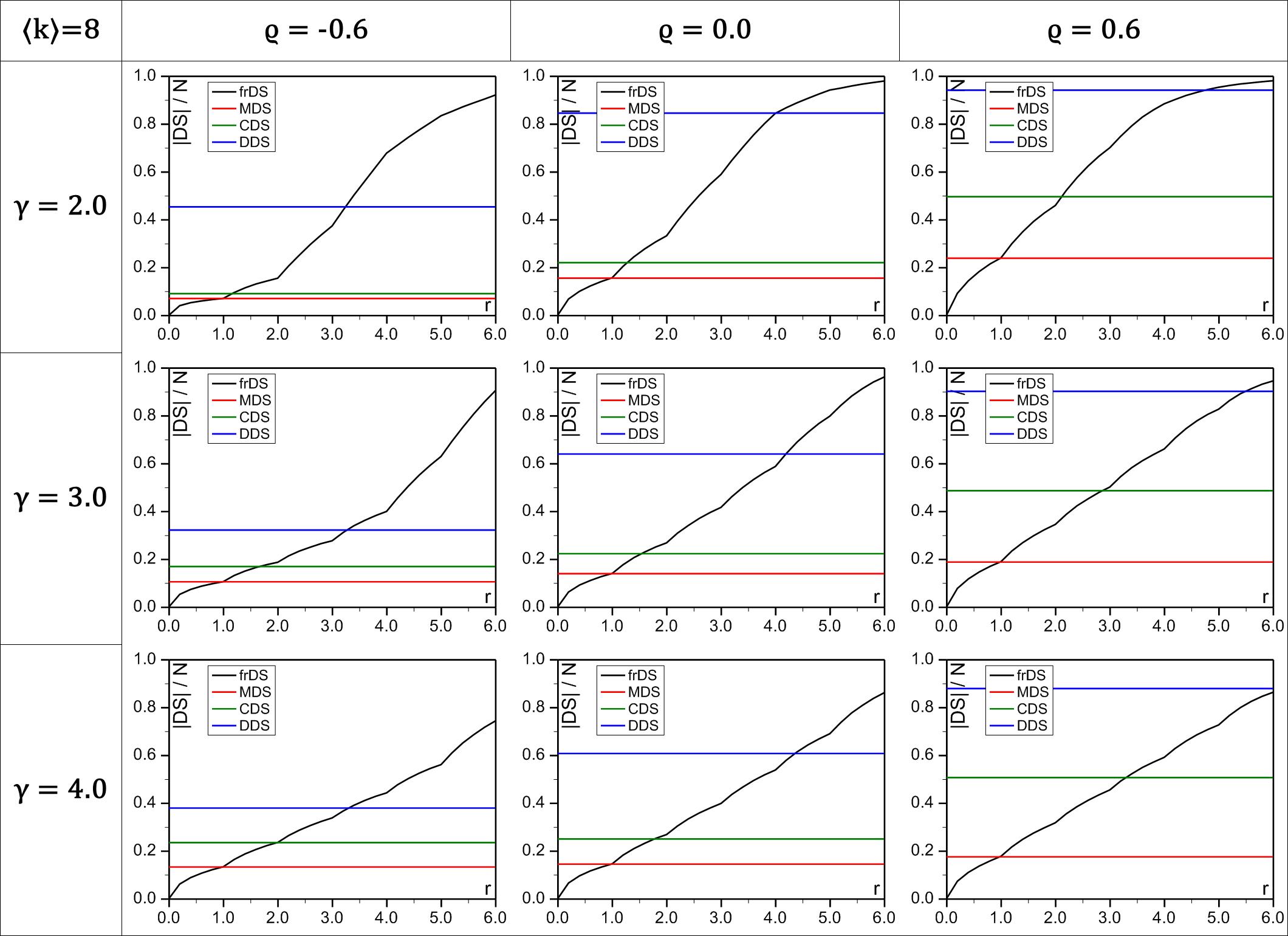}
\caption{Size of frDS as a function of domination redundancy, at various power-law degree exponents and Spearman's $\rho$ values, in synthetic networks with $N=5000$. The sizes of MDS, CDS, and DDS are shown for comparison.}
\label{fig-s6}
\end{figure}

\clearpage

\begin{figure}[h!]
\begin{center}
\includegraphics[width=\textwidth]{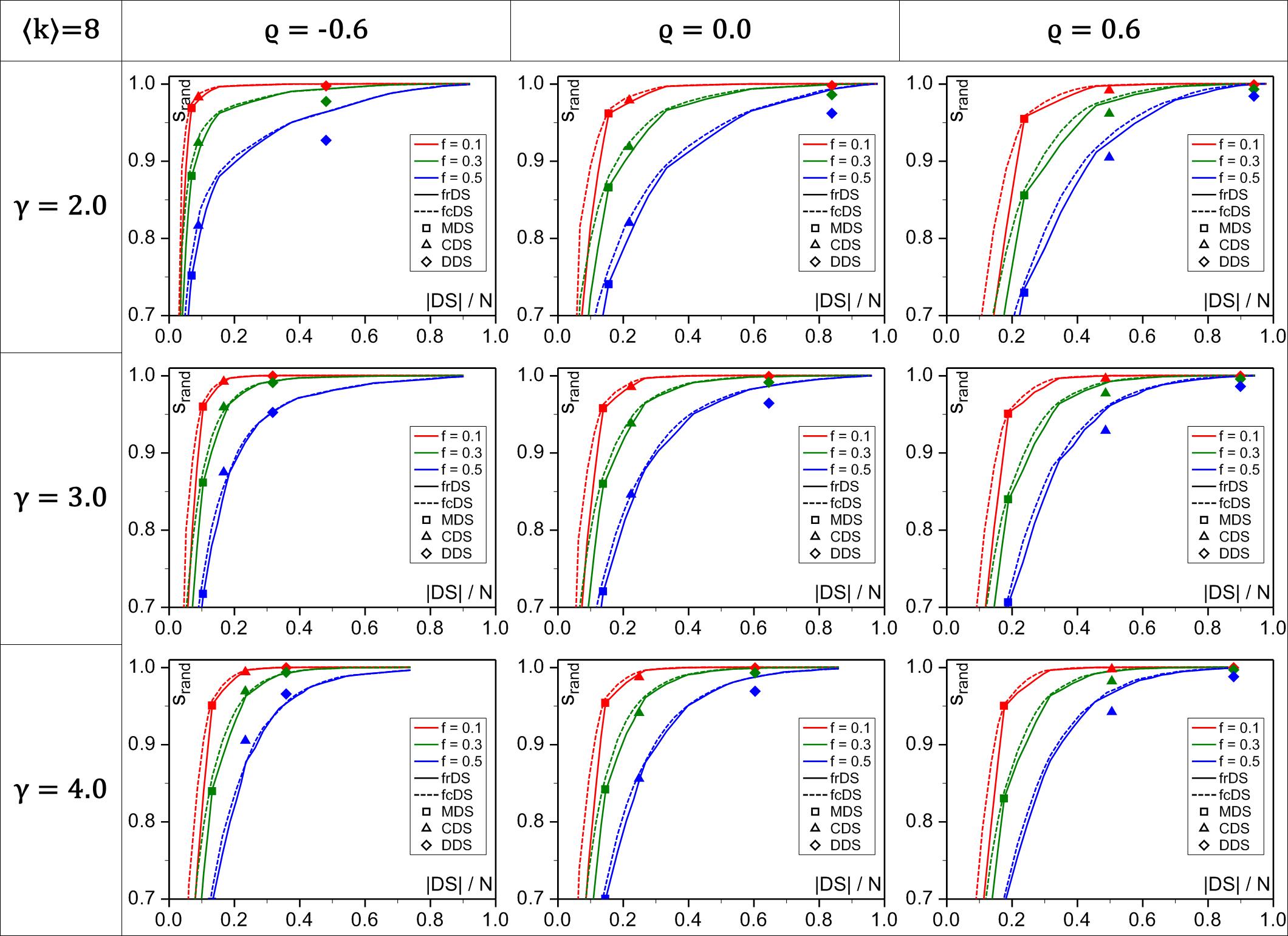}
\caption{Stability of frDS and fcDS against random damage, as a function of dominating set size (cost), at various power-law degree exponents and Spearman's $\rho$ values, in synthetic networks with $N=5000$. The stabilities of MDS, CDS, and DDS are shown for comparison at their corresponding set sizes. Black legend symbols refer to the shape only, colors refer to damage fractions.}
\label{fig-s7}
\end{center}
\end{figure}

\begin{figure}[h!]
\begin{center}
\includegraphics[width=\textwidth]{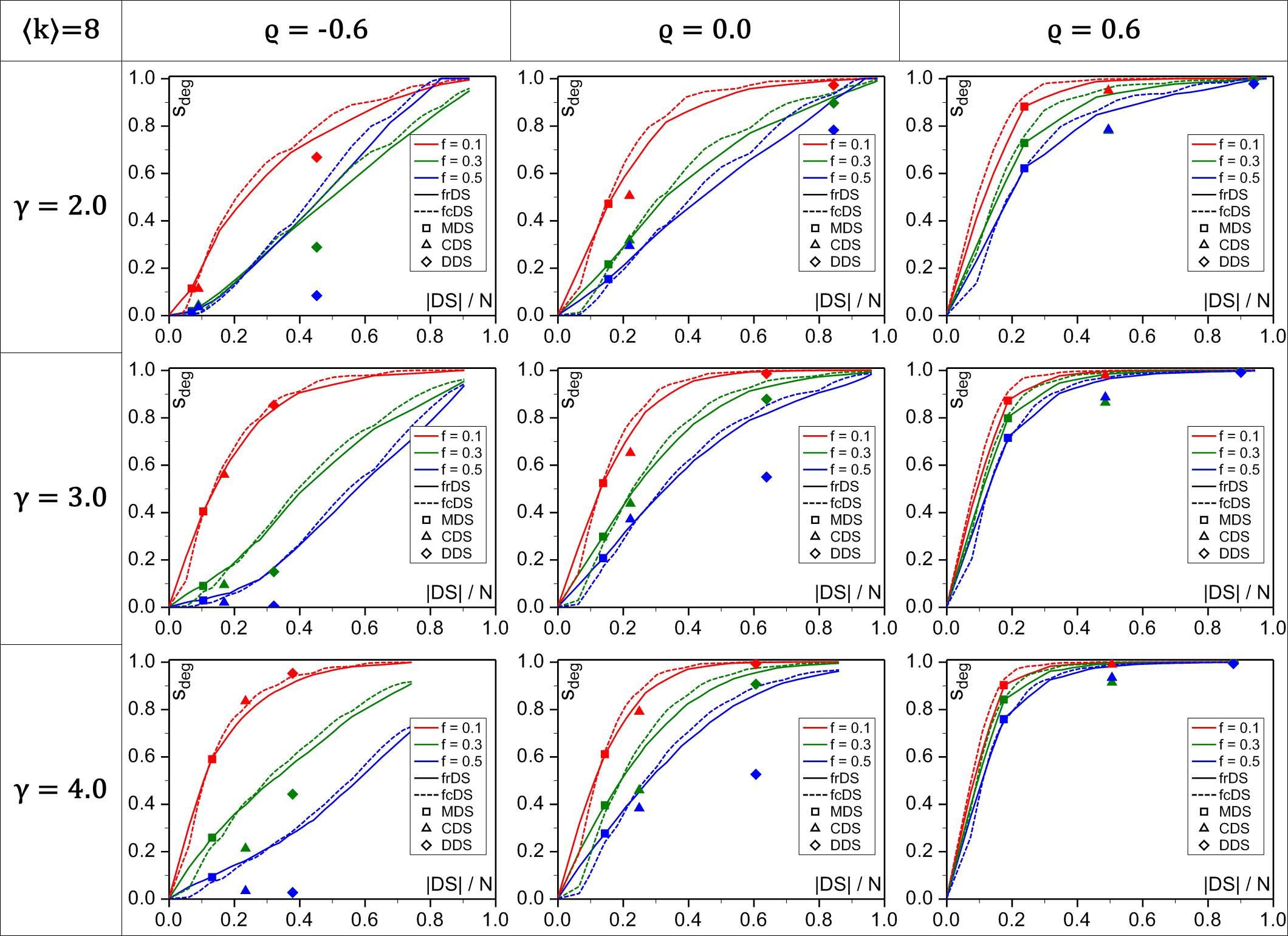}
\end{center}
\caption{Stability of frDS and fcDS against targeted attack, as a function of dominating set size (cost), at various power-law degree exponents and Spearman's $\rho$ values, in synthetic networks with $N=5000$. The stabilities of MDS, CDS, and DDS are shown for comparison at their corresponding set sizes. Black legend symbols refer to the shape only, colors refer to damage fractions.}
\label{fig-s8}
\end{figure}

\clearpage

\begin{figure}[h!]
\includegraphics[width=\textwidth]{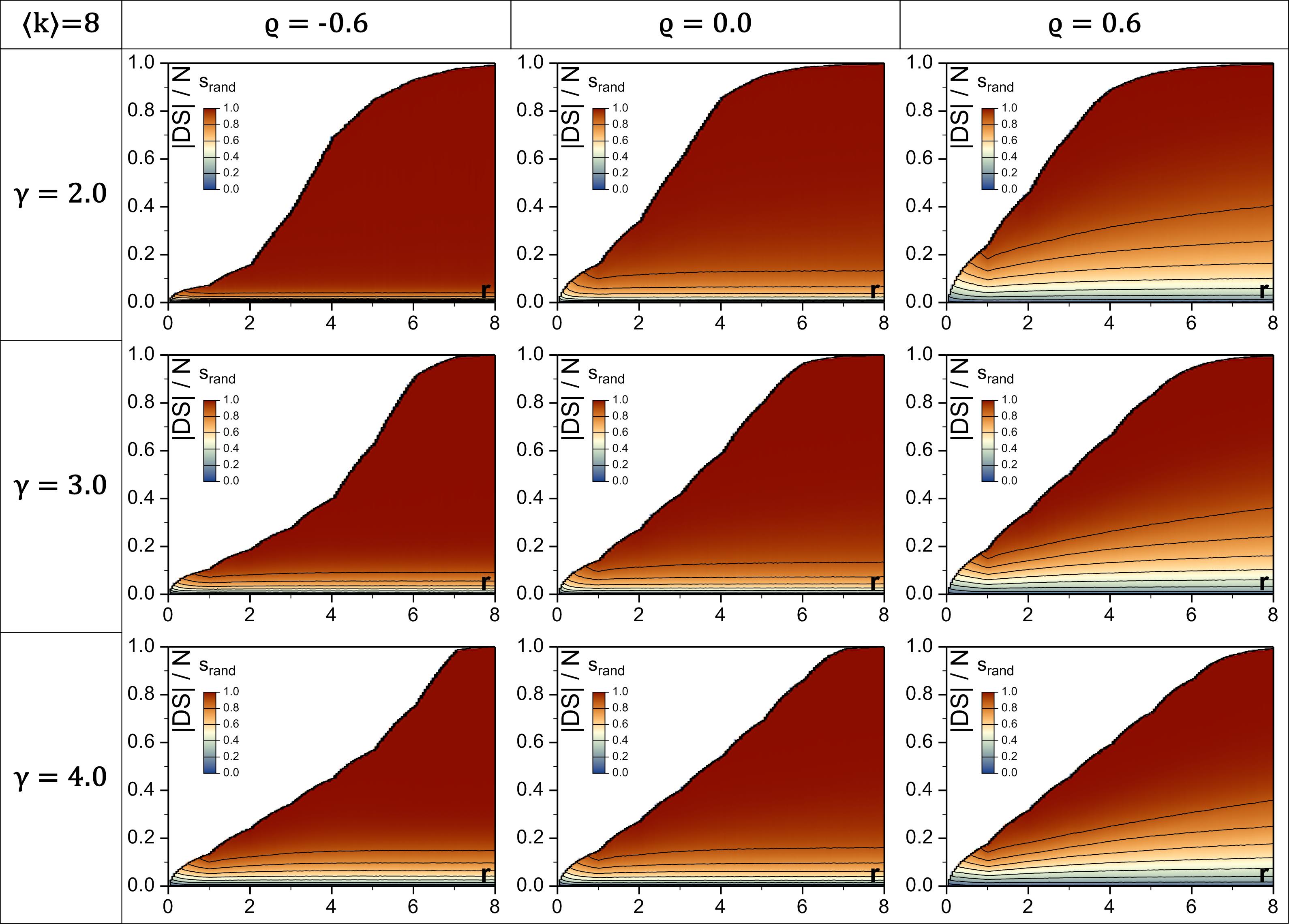}
\caption{Stability of partial frDS against random damage, as a function of redundancy level and dominating set size, at various power-law degree exponents and Spearman's $\rho$ values. Synthetic networks, $N=5000$, damage fraction $f=0.1$.}
\label{fig-s9}
\end{figure}

\begin{figure}[h!]
\includegraphics[width=\textwidth]{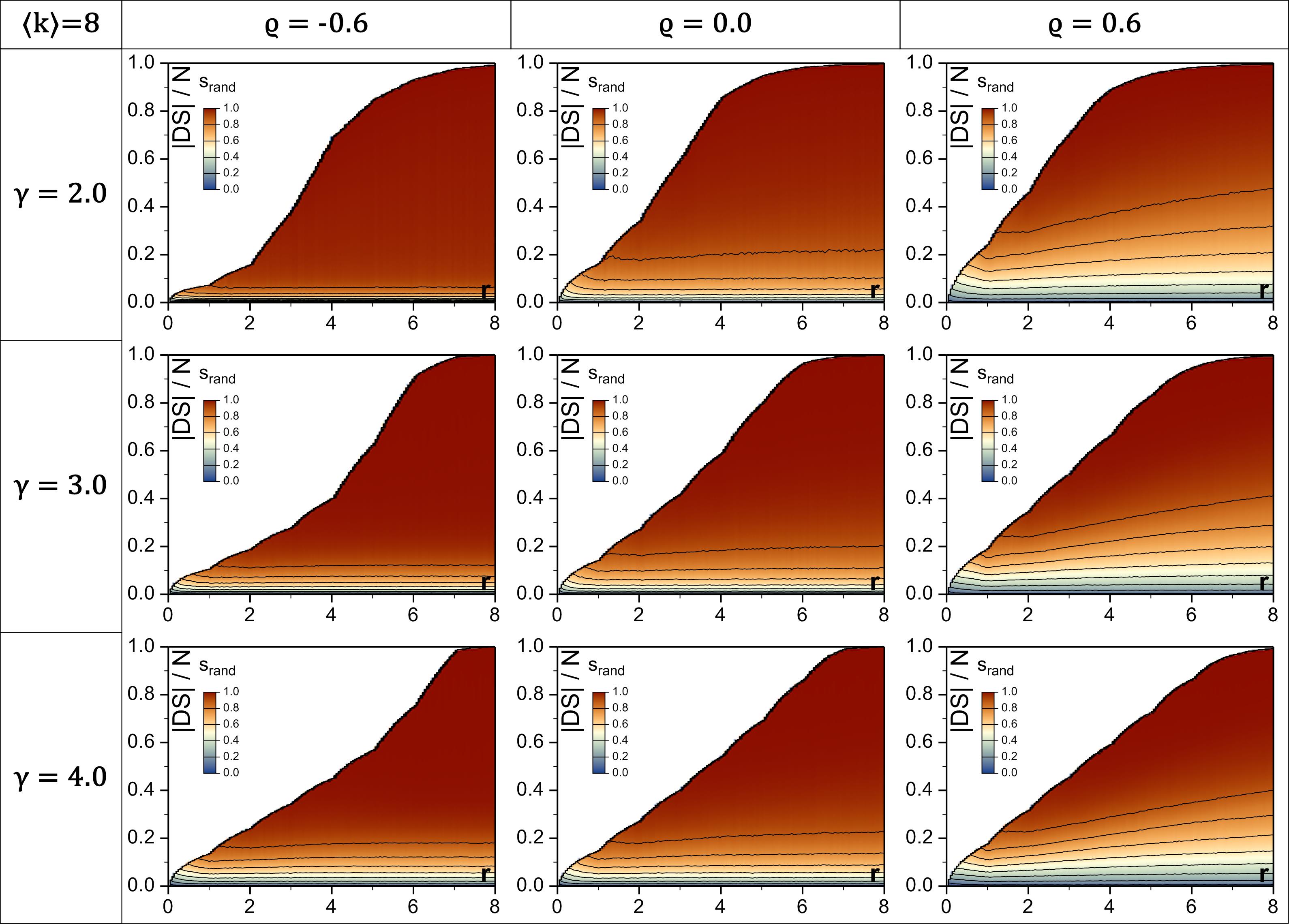}
\caption{Stability of partial frDS against random damage, as a function of redundancy level and dominating set size, at various power-law degree exponents and Spearman's $\rho$ values. Synthetic networks, $N=5000$, damage fraction $f=0.3$.}
\label{fig-s10}
\end{figure}

\begin{figure}[h!]
\includegraphics[width=\textwidth]{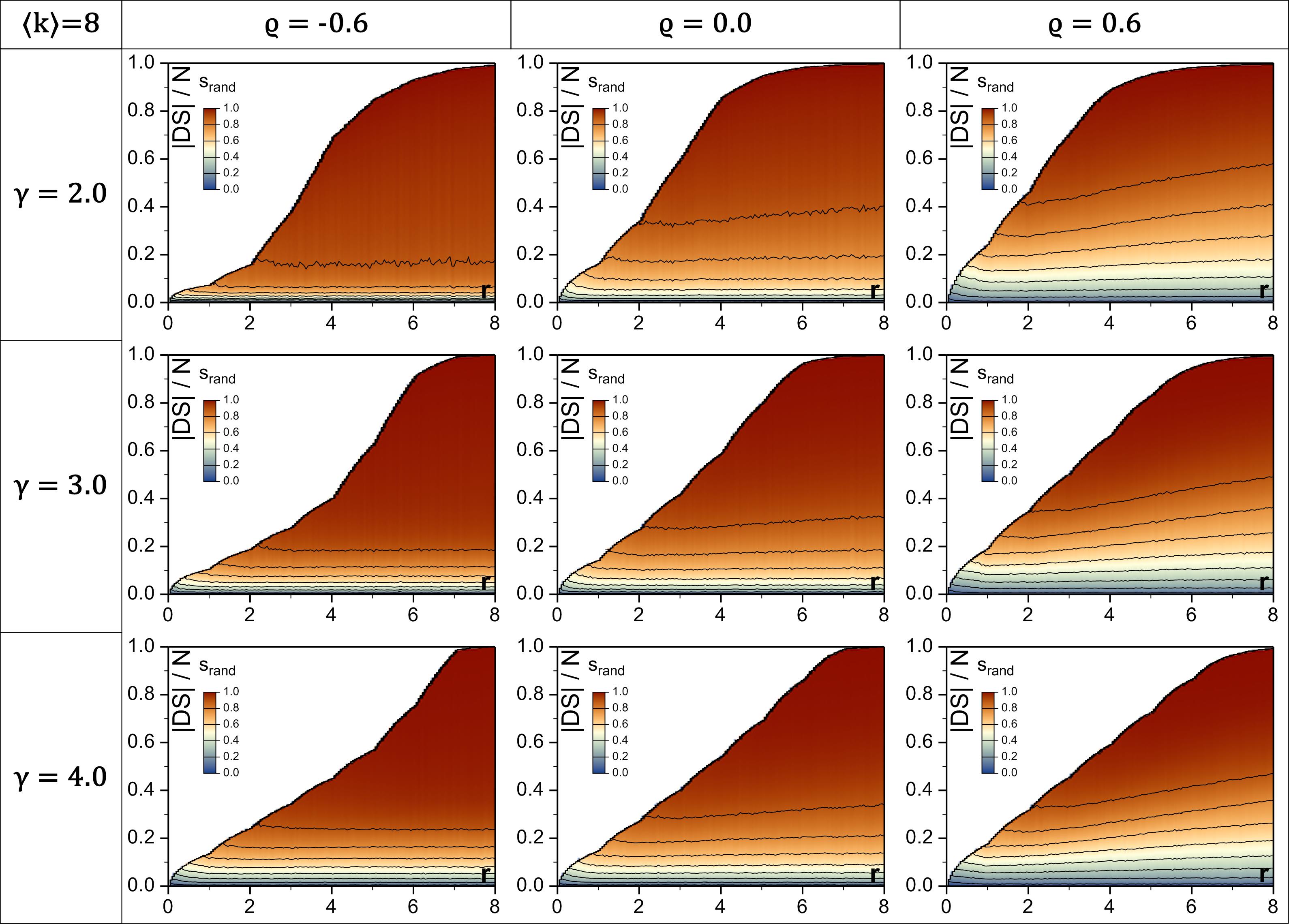}
\caption{Stability of partial frDS against random damage, as a function of redundancy level and dominating set size, at various power-law degree exponents and Spearman's $\rho$ values. Synthetic networks, $N=5000$, damage fraction $f=0.5$.}
\label{fig-s11}
\end{figure}

\begin{figure}[h!]
\includegraphics[width=\textwidth]{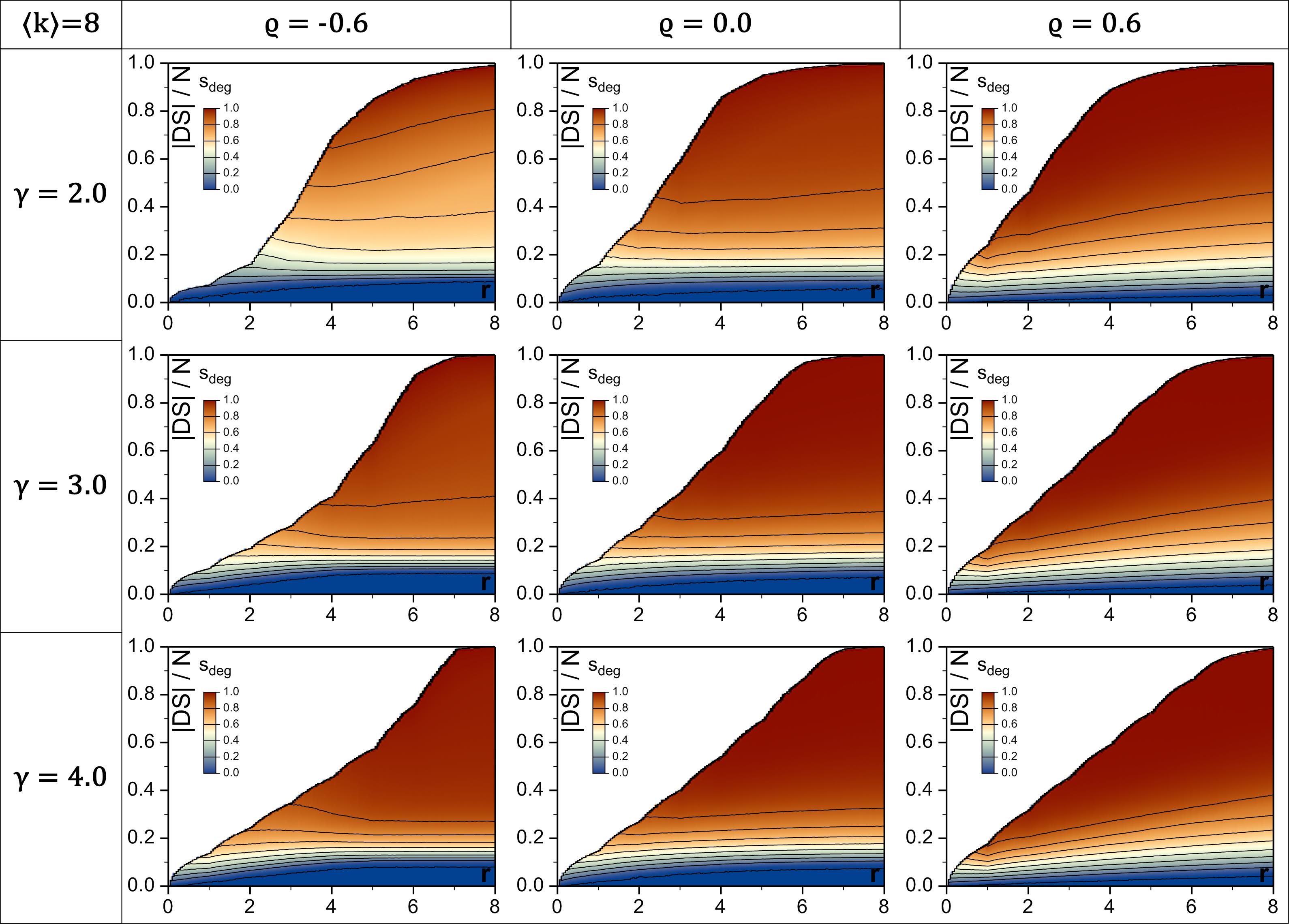}
\caption{Stability of partial frDS against targeted attack, as a function of redundancy level and dominating set size, at various power-law degree exponents and Spearman's $\rho$ values. Synthetic networks, $N=5000$, damage fraction $f=0.1$.}
\label{fig-s12}
\end{figure}

\begin{figure}[h!]
\includegraphics[width=\textwidth]{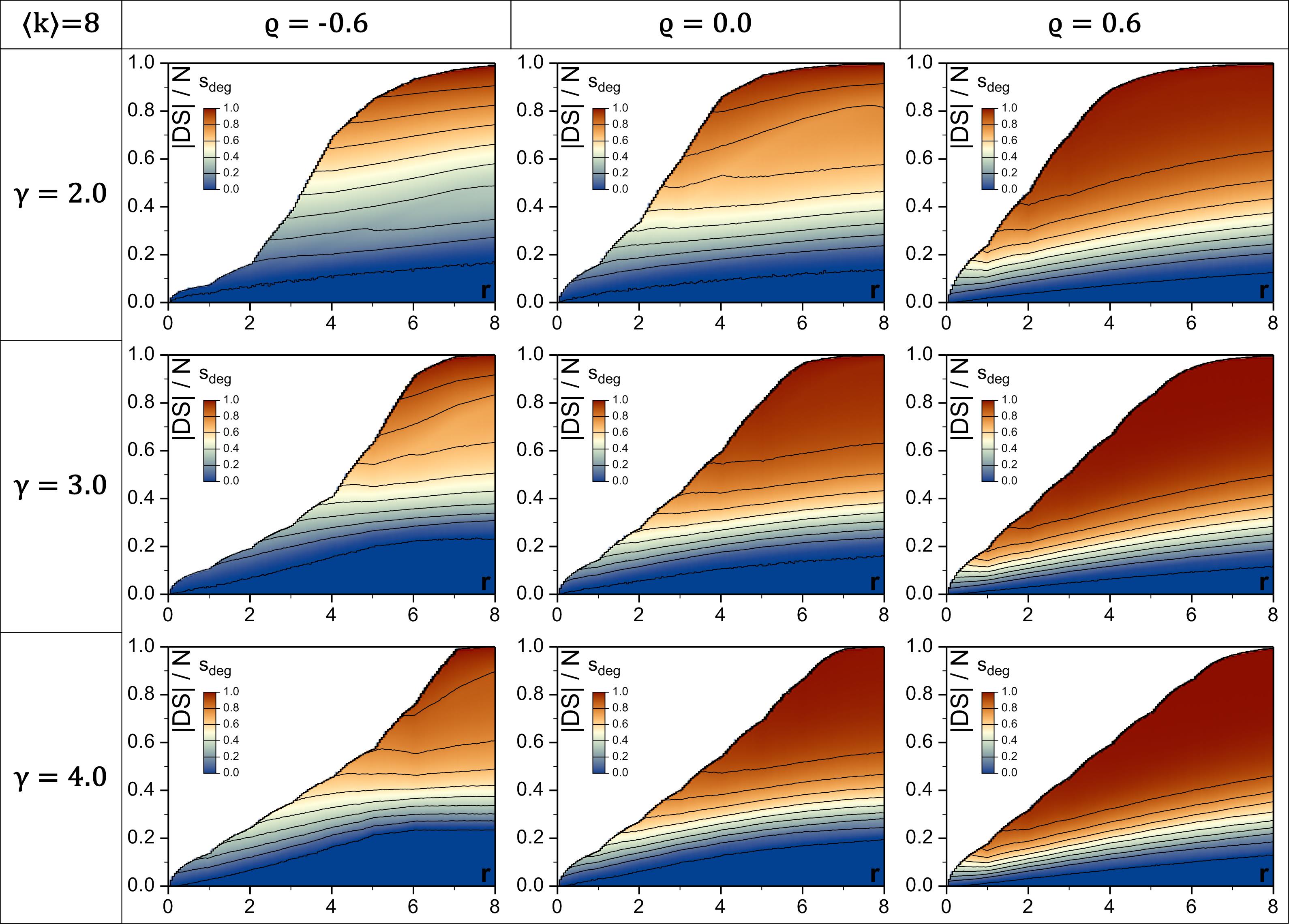}
\caption{Stability of partial frDS against targeted attack, as a function of redundancy level and dominating set size, at various power-law degree exponents and Spearman's $\rho$ values. Synthetic networks, $N=5000$, damage fraction $f=0.3$.}
\label{fig-s13}
\end{figure}

\begin{figure}[h!]
\includegraphics[width=\textwidth]{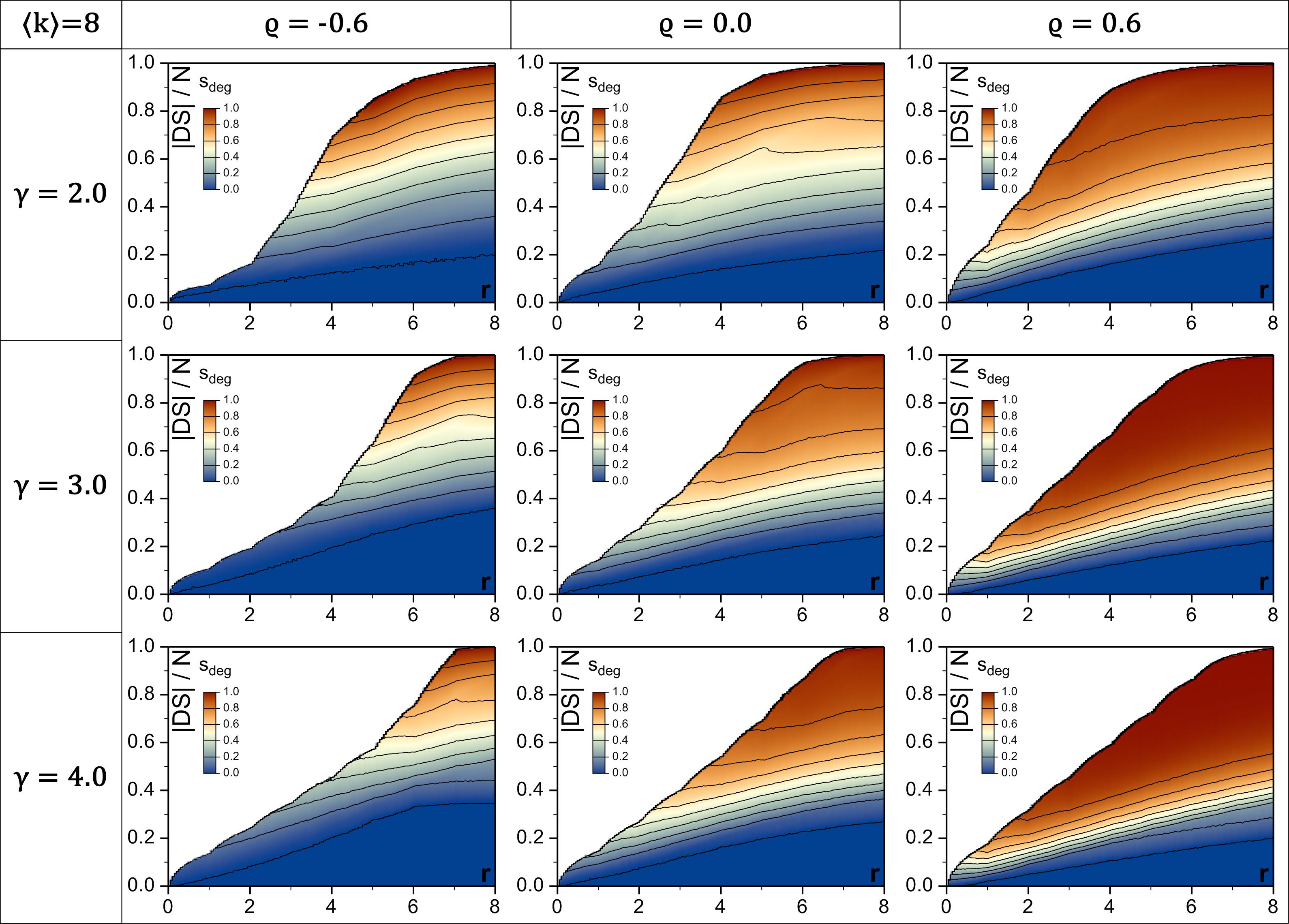}
\caption{Stability of partial frDS against targeted attack, as a function of redundancy level and dominating set size, at various power-law degree exponents and Spearman's $\rho$ values. Synthetic networks, $N=5000$, damage fraction $f=0.5$.}
\label{fig-s14}
\end{figure}

\clearpage

\begin{figure}[h!]
\begin{center}
\includegraphics[width=3in]{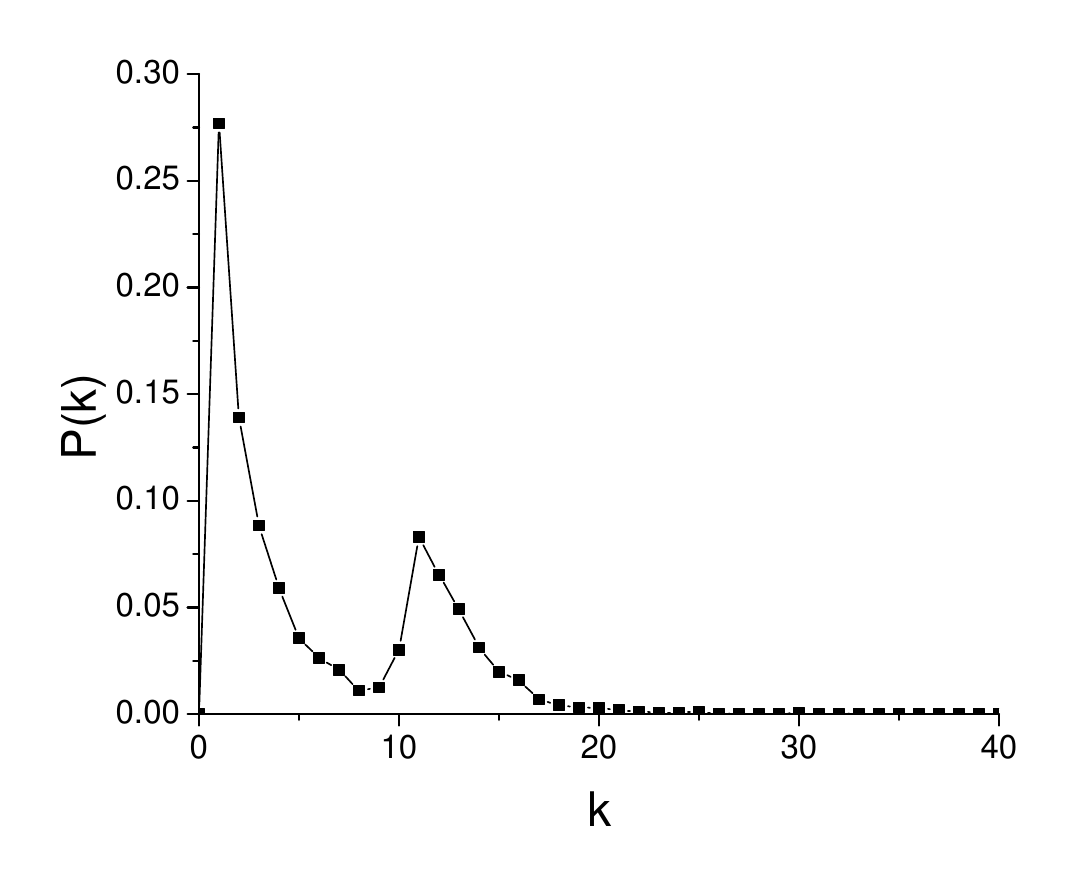}
\caption{Degree distribution of Gnutella08 network \cite{s_stanford} on linear scale.}
\label{fig-dd-gnutella-linlin}
\end{center}
\end{figure}

\begin{figure}[h!]
\begin{center}
\includegraphics[width=3in]{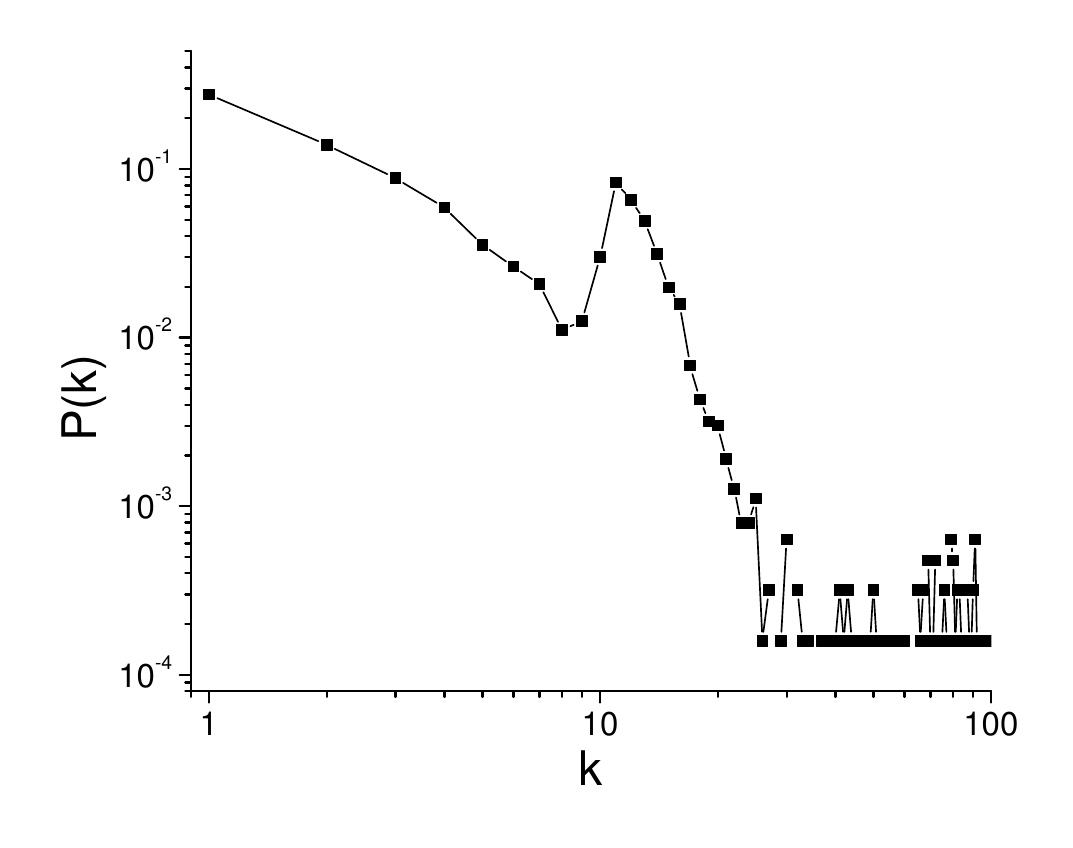}
\caption{Degree distribution of Gnutella08 network \cite{s_stanford} on double-logarithmic scale.}
\label{fig-dd-gnutella-loglog}
\end{center}
\end{figure}

\begin{figure}[h!]
\begin{center}
\includegraphics[width=3in]{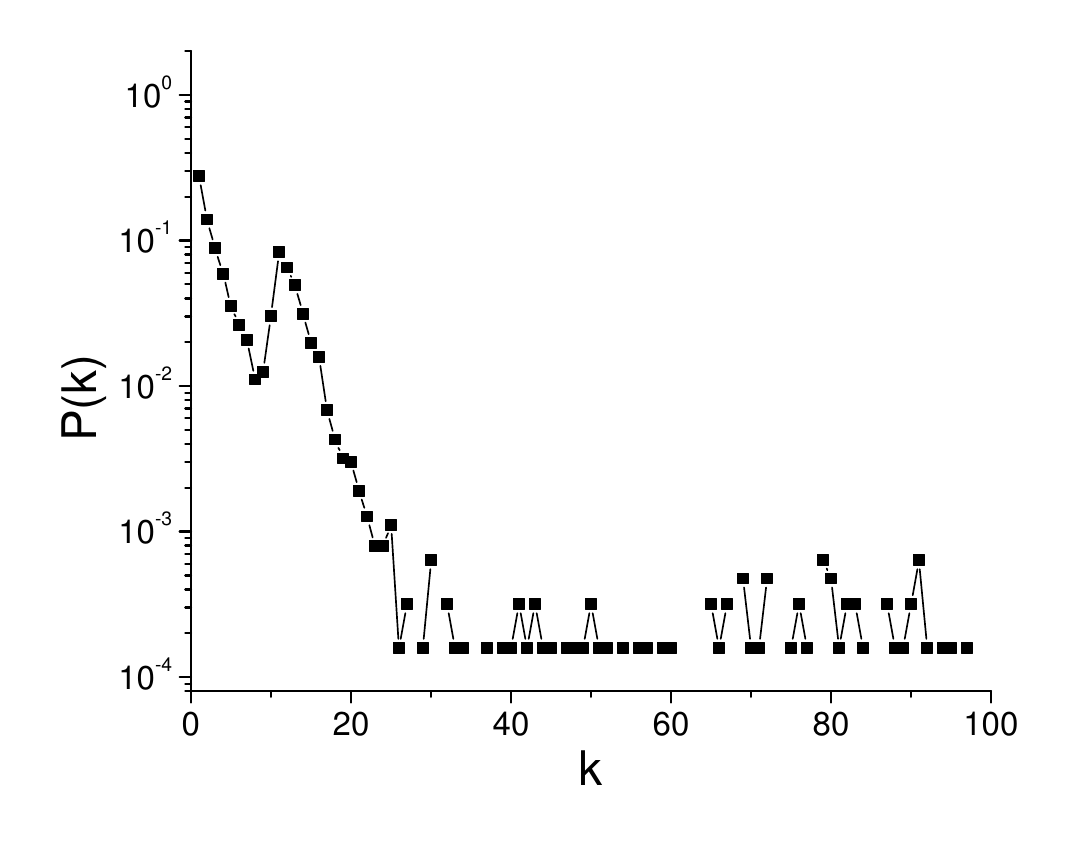}
\caption{Degree distribution of Gnutella08 network \cite{s_stanford} on log-linear scale.}
\label{fig-dd-gnutella-loglin}
\end{center}
\end{figure}

\clearpage

\begin{figure}[h!]
\begin{center}
\includegraphics[width=3in]{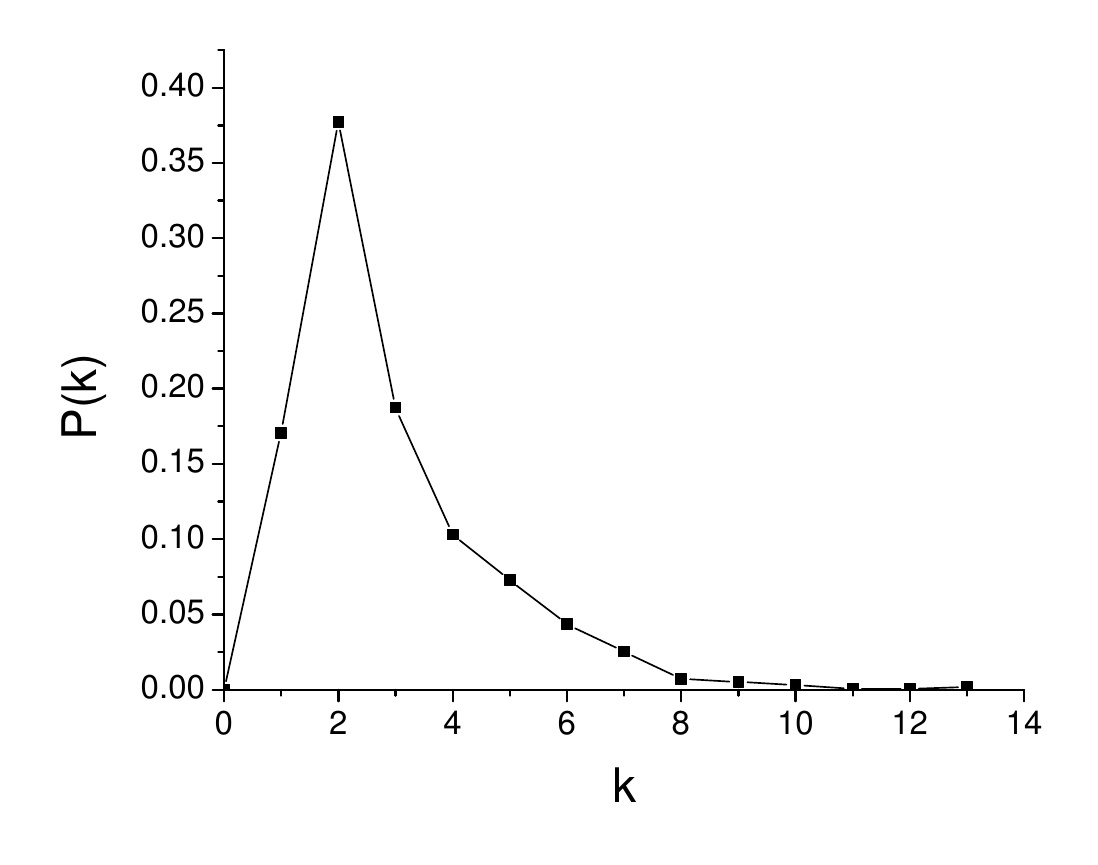}
\caption{Degree distribution of ENTSO-E powergrid \cite{s_powergrid1,s_powergrid2} on linear scale.}
\label{fig-dd-pgrid-linlin}
\end{center}
\end{figure}

\begin{figure}[h!]
\begin{center}
\includegraphics[width=3in]{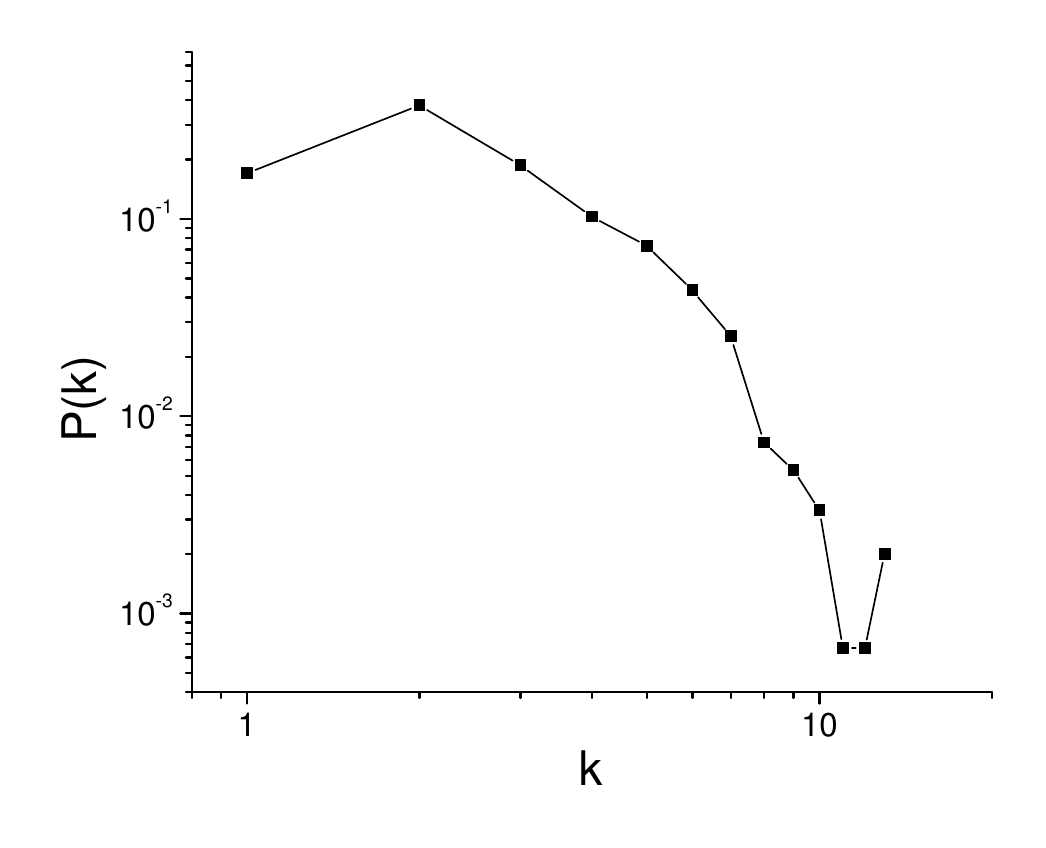}
\caption{Degree distribution of ENTSO-E powergrid \cite{s_powergrid1,s_powergrid2} on double-logarithmic scale.}
\label{fig-dd-pgrid-loglog}
\end{center}
\end{figure}

\begin{figure}[h!]
\begin{center}
\includegraphics[width=3in]{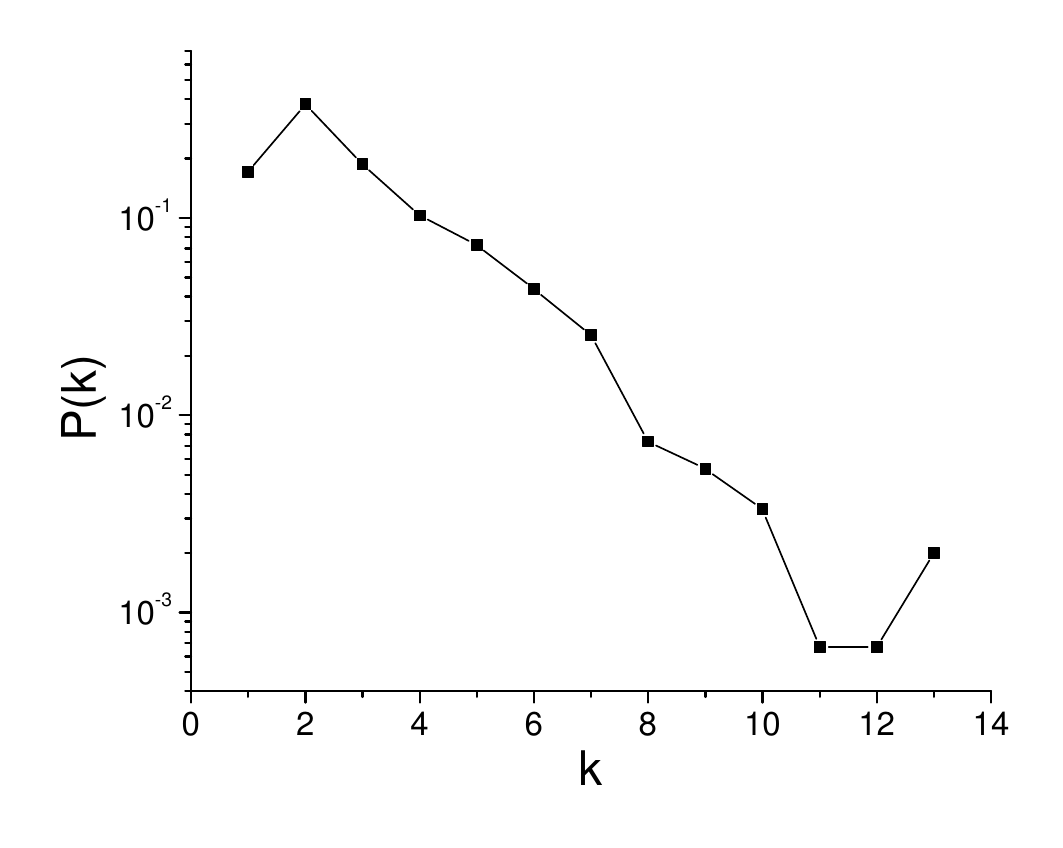}
\caption{Degree distribution of ENTSO-E powergrid \cite{s_powergrid1,s_powergrid2} on log-linear scale.}
\label{fig-dd-pgrid-loglin}
\end{center}
\end{figure}

\clearpage

\begin{figure}[h!]
\begin{center}
\includegraphics[width=5in]{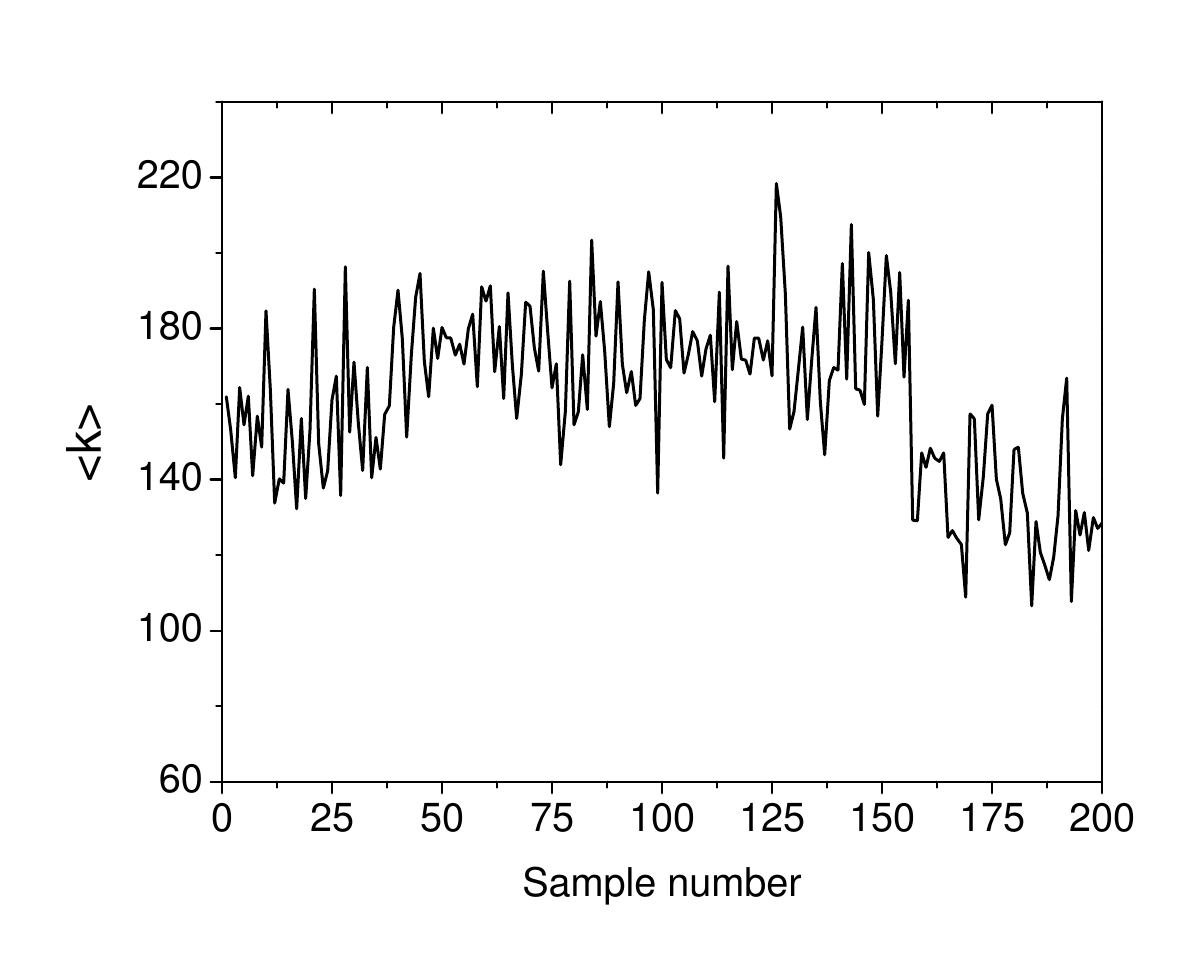}
\caption{Average degree in brain graphs. See Supplementary Table~\ref{table-samples} for sample numbers.}
\label{fig-kavg-sample}
\end{center}
\end{figure}

\begin{figure}[h!]
\begin{center}
\includegraphics[width=4.6in]{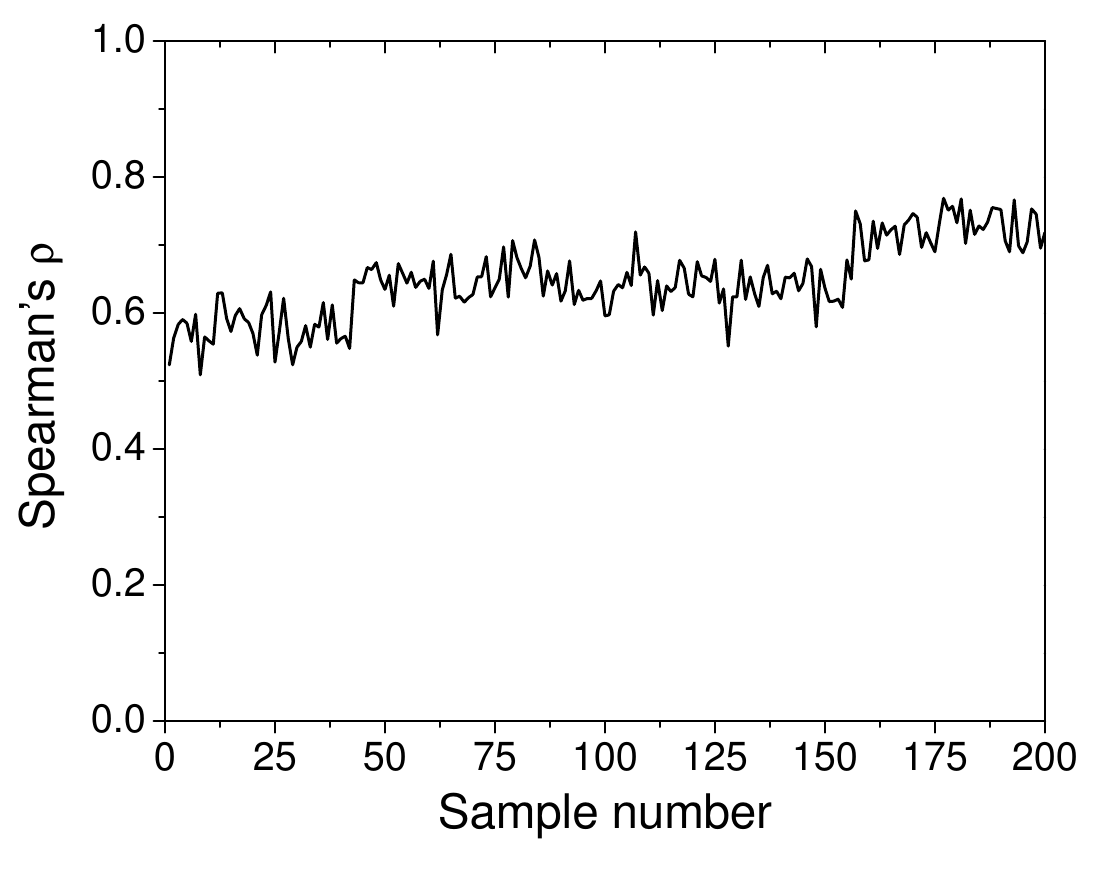}
\caption{Assortativity of brain graphs. See Supplementary Table~\ref{table-samples} for sample numbers.}
\label{fig-rho-sample}
\end{center}
\end{figure}

\begin{figure}[h!]
\begin{center}
\includegraphics[width=5in]{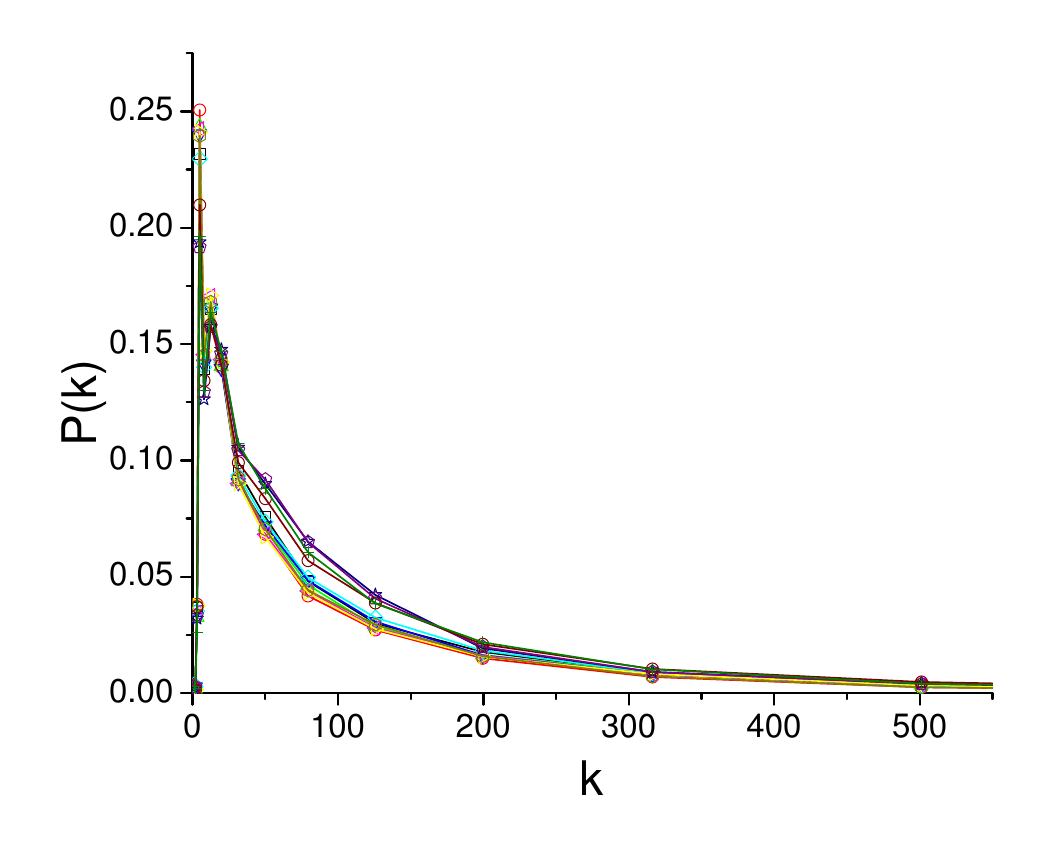}
\caption{Degree distributions of $12$ randomly picked brain graphs on linear scale.}
\label{fig-dd-brain-linlin}
\end{center}
\end{figure}

\begin{figure}[h!]
\begin{center}
\includegraphics[width=5in]{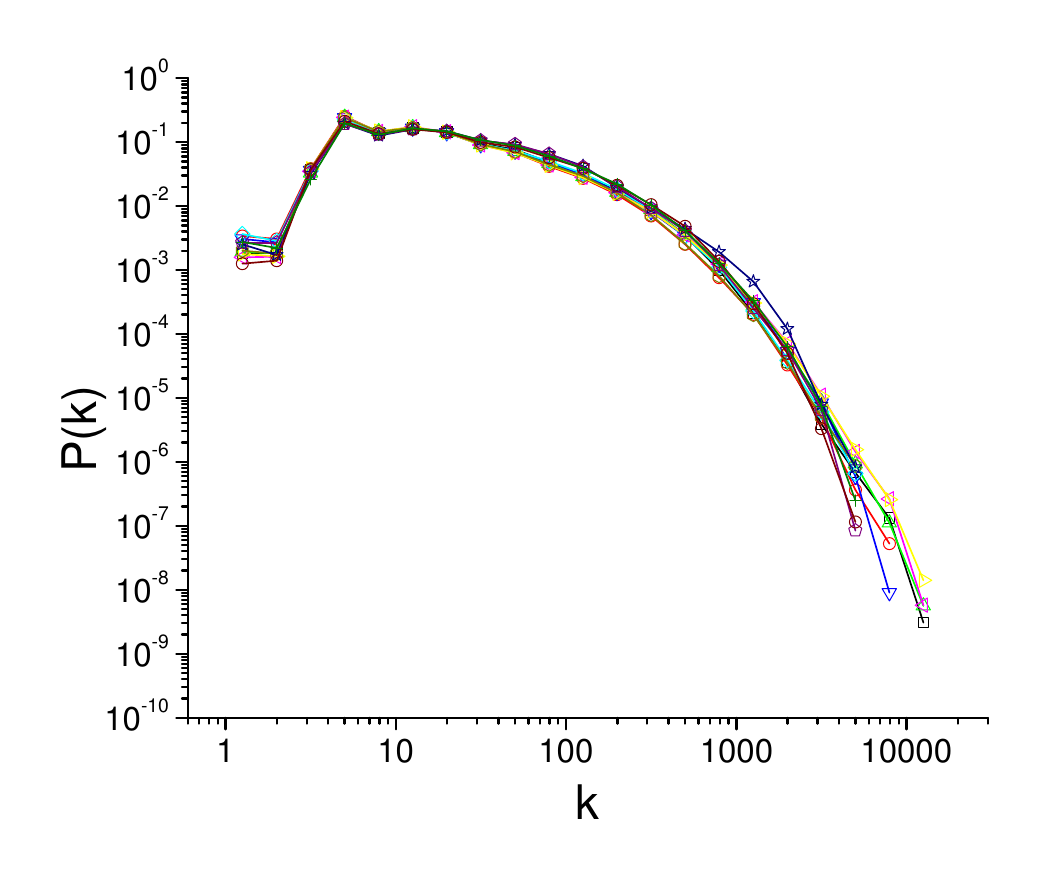}
\caption{Degree distributions of $12$ randomly picked brain graphs on double-logarithmic scale.}
\label{fig-dd-brain-loglog}
\end{center}
\end{figure}

\begin{figure}[h!]
\begin{center}
\includegraphics[width=5in]{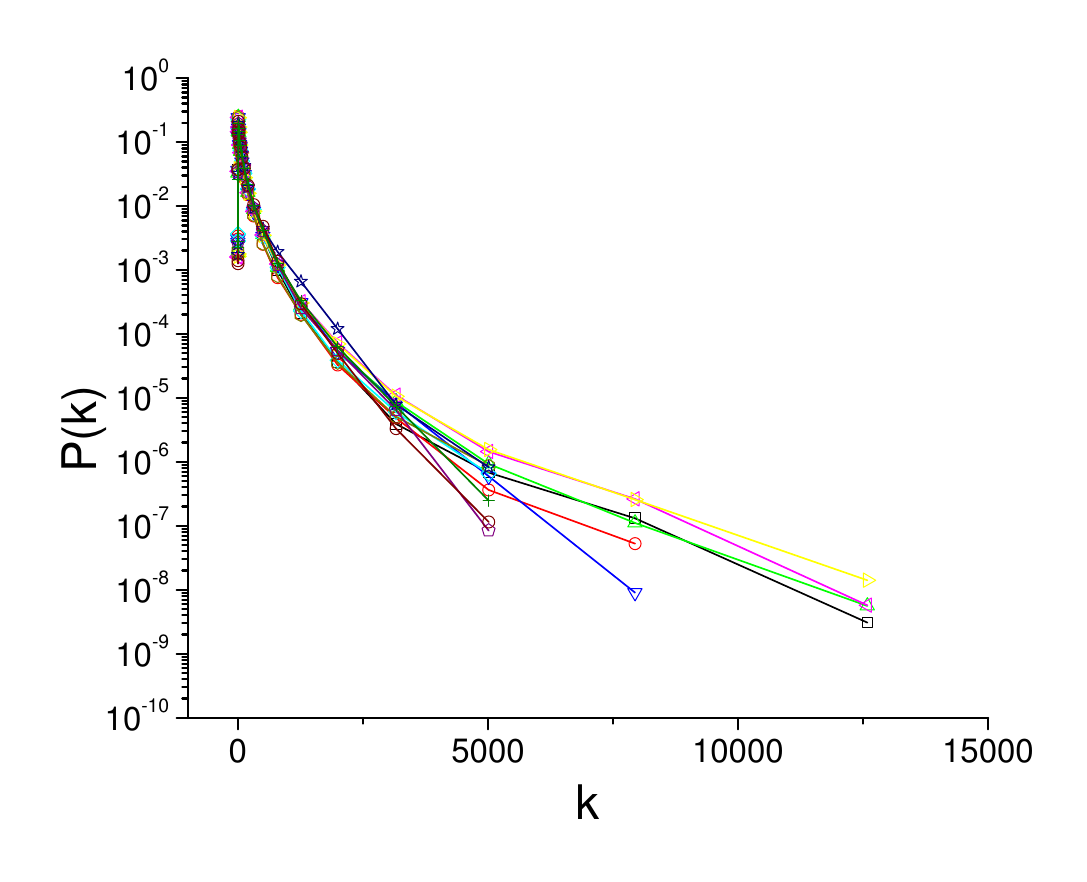}
\caption{Degree distributions of $12$ randomly picked brain graphs on log-linear scale.}
\label{fig-dd-brain-loglin}
\end{center}
\end{figure}

\begin{figure}[h!]
\begin{center}
\includegraphics[width=5in]{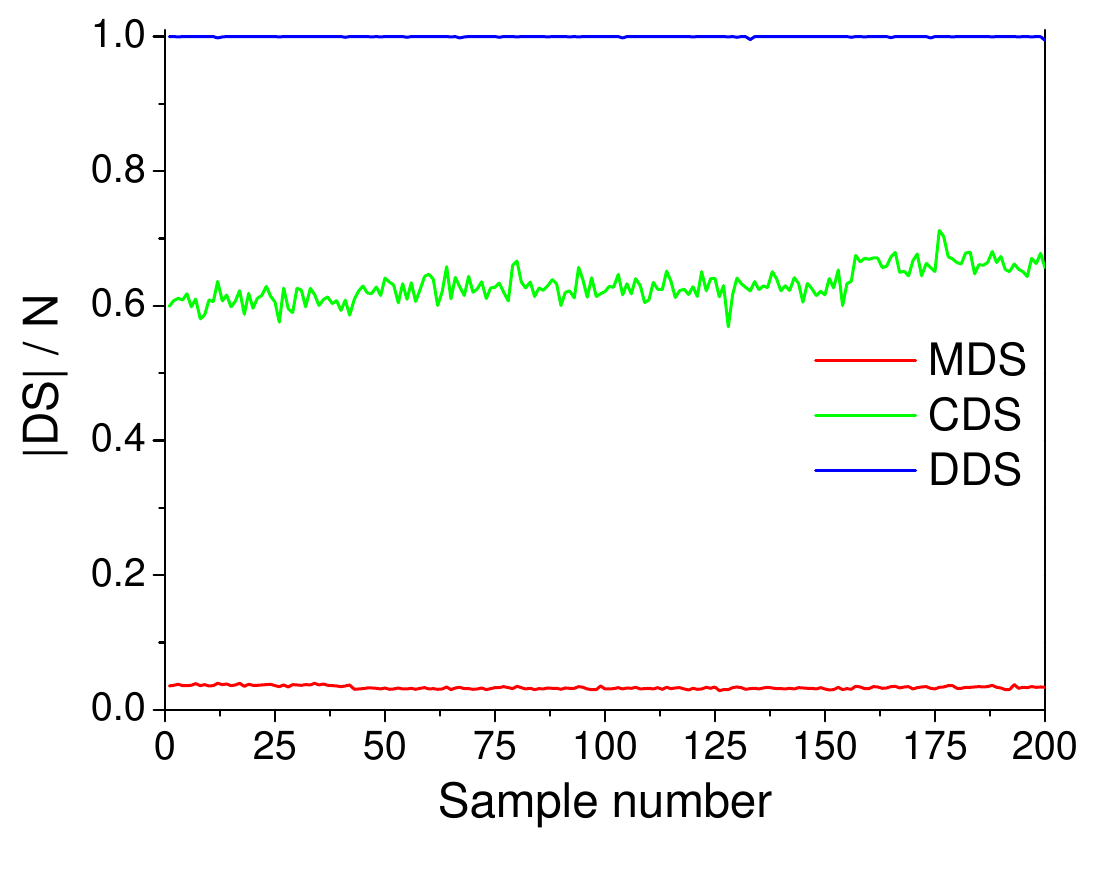}
\caption{Comparison of dominating set sizes in brain graphs. See Supplementary Table~\ref{table-samples} for sample numbers.}
\label{fig-ds-sample}
\end{center}
\end{figure}

\clearpage

\begin{center}
\small
\begin{longtable}{| c | c | c | c |}
\hline
sample\# & graph & sample\# & graph \\ \hline
1 & KKI-21\_KKI2009-01\_big\_graph\_w\_inv & 51 & MRN114\_M87114047\_big\_graph\_w\_inv \\ \hline
2 & KKI-21\_KKI2009-02\_big\_graph\_w\_inv & 52 & MRN114\_M87114064\_big\_graph\_w\_inv \\ \hline
3 & KKI-21\_KKI2009-03\_big\_graph\_w\_inv & 53 & MRN114\_M87115498\_big\_graph\_w\_inv \\ \hline
4 & KKI-21\_KKI2009-04\_big\_graph\_w\_inv & 54 & MRN114\_M87115517\_big\_graph\_w\_inv \\ \hline
5 & KKI-21\_KKI2009-05\_big\_graph\_w\_inv & 55 & MRN114\_M87117119\_big\_graph\_w\_inv \\ \hline
6 & KKI-21\_KKI2009-06\_big\_graph\_w\_inv & 56 & MRN114\_M87117167\_big\_graph\_w\_inv \\ \hline
7 & KKI-21\_KKI2009-07\_big\_graph\_w\_inv & 57 & MRN114\_M87120962\_big\_graph\_w\_inv \\ \hline
8 & KKI-21\_KKI2009-08\_big\_graph\_w\_inv & 58 & MRN114\_M87121943\_big\_graph\_w\_inv \\ \hline
9 & KKI-21\_KKI2009-09\_big\_graph\_w\_inv & 59 & MRN114\_M87121956\_big\_graph\_w\_inv \\ \hline
10 & KKI-21\_KKI2009-10\_big\_graph\_w\_inv & 60 & MRN114\_M87122092\_big\_graph\_w\_inv \\ \hline
11 & KKI-21\_KKI2009-11\_big\_graph\_w\_inv & 61 & MRN114\_M87123042\_big\_graph\_w\_inv \\ \hline
12 & KKI-21\_KKI2009-12\_big\_graph\_w\_inv & 62 & MRN114\_M87123449\_big\_graph\_w\_inv \\ \hline
13 & KKI-21\_KKI2009-13\_big\_graph\_w\_inv & 63 & MRN114\_M87123913\_big\_graph\_w\_inv \\ \hline
14 & KKI-21\_KKI2009-14\_big\_graph\_w\_inv & 64 & MRN114\_M87124633\_big\_graph\_w\_inv \\ \hline
15 & KKI-21\_KKI2009-15\_big\_graph\_w\_inv & 65 & MRN114\_M87124781\_big\_graph\_w\_inv \\ \hline
16 & KKI-21\_KKI2009-16\_big\_graph\_w\_inv & 66 & MRN114\_M87124827\_big\_graph\_w\_inv \\ \hline
17 & KKI-21\_KKI2009-17\_big\_graph\_w\_inv & 67 & MRN114\_M87125134\_big\_graph\_w\_inv \\ \hline
18 & KKI-21\_KKI2009-18\_big\_graph\_w\_inv & 68 & MRN114\_M87128444\_big\_graph\_w\_inv \\ \hline
19 & KKI-21\_KKI2009-19\_big\_graph\_w\_inv & 69 & MRN114\_M87129719\_big\_graph\_w\_inv \\ \hline
20 & KKI-21\_KKI2009-20\_big\_graph\_w\_inv & 70 & MRN114\_M87129789\_big\_graph\_w\_inv \\ \hline
21 & KKI-21\_KKI2009-21\_big\_graph\_w\_inv & 71 & MRN114\_M87131806\_big\_graph\_w\_inv \\ \hline
22 & KKI-21\_KKI2009-22\_big\_graph\_w\_inv & 72 & MRN114\_M87134068\_big\_graph\_w\_inv \\ \hline
23 & KKI-21\_KKI2009-23\_big\_graph\_w\_inv & 73 & MRN114\_M87135647\_big\_graph\_w\_inv \\ \hline
24 & KKI-21\_KKI2009-24\_big\_graph\_w\_inv & 74 & MRN114\_M87136332\_big\_graph\_w\_inv \\ \hline
25 & KKI-21\_KKI2009-25\_big\_graph\_w\_inv & 75 & MRN114\_M87136832\_big\_graph\_w\_inv \\ \hline
26 & KKI-21\_KKI2009-26\_big\_graph\_w\_inv & 76 & MRN114\_M87139021\_big\_graph\_w\_inv \\ \hline
27 & KKI-21\_KKI2009-27\_big\_graph\_w\_inv & 77 & MRN114\_M87139257\_big\_graph\_w\_inv \\ \hline
28 & KKI-21\_KKI2009-28\_big\_graph\_w\_inv & 78 & MRN114\_M87141220\_big\_graph\_w\_inv \\ \hline
29 & KKI-21\_KKI2009-29\_big\_graph\_w\_inv & 79 & MRN114\_M87141664\_big\_graph\_w\_inv \\ \hline
30 & KKI-21\_KKI2009-30\_big\_graph\_w\_inv & 80 & MRN114\_M87141793\_big\_graph\_w\_inv \\ \hline
31 & KKI-21\_KKI2009-31\_big\_graph\_w\_inv & 81 & MRN114\_M87141858\_big\_graph\_w\_inv \\ \hline
32 & KKI-21\_KKI2009-32\_big\_graph\_w\_inv & 82 & MRN114\_M87141906\_big\_graph\_w\_inv \\ \hline
33 & KKI-21\_KKI2009-33\_big\_graph\_w\_inv & 83 & MRN114\_M87141949\_big\_graph\_w\_inv \\ \hline
34 & KKI-21\_KKI2009-34\_big\_graph\_w\_inv & 84 & MRN114\_M87142764\_big\_graph\_w\_inv \\ \hline
35 & KKI-21\_KKI2009-35\_big\_graph\_w\_inv & 85 & MRN114\_M87143273\_big\_graph\_w\_inv \\ \hline
36 & KKI-21\_KKI2009-36\_big\_graph\_w\_inv & 86 & MRN114\_M87144889\_big\_graph\_w\_inv \\ \hline
37 & KKI-21\_KKI2009-37\_big\_graph\_w\_inv & 87 & MRN114\_M87144896\_big\_graph\_w\_inv \\ \hline
38 & KKI-21\_KKI2009-38\_big\_graph\_w\_inv & 88 & MRN114\_M87145479\_big\_graph\_w\_inv \\ \hline
39 & KKI-21\_KKI2009-39\_big\_graph\_w\_inv & 89 & MRN114\_M87145575\_big\_graph\_w\_inv \\ \hline
40 & KKI-21\_KKI2009-40\_big\_graph\_w\_inv & 90 & MRN114\_M87146520\_big\_graph\_w\_inv \\ \hline
41 & KKI-21\_KKI2009-41\_big\_graph\_w\_inv & 91 & MRN114\_M87146993\_big\_graph\_w\_inv \\ \hline
42 & KKI-21\_KKI2009-42\_big\_graph\_w\_inv & 92 & MRN114\_M87147006\_big\_graph\_w\_inv \\ \hline
43 & MRN114\_M87102217\_big\_graph\_w\_inv & 93 & MRN114\_M87148745\_big\_graph\_w\_inv \\ \hline
44 & MRN114\_M87102806\_big\_graph\_w\_inv & 94 & MRN114\_M87149014\_big\_graph\_w\_inv \\ \hline
45 & MRN114\_M87103074\_big\_graph\_w\_inv & 95 & MRN114\_M87149025\_big\_graph\_w\_inv \\ \hline
46 & MRN114\_M87105476\_big\_graph\_w\_inv & 96 & MRN114\_M87150194\_big\_graph\_w\_inv \\ \hline
47 & MRN114\_M87107085\_big\_graph\_w\_inv & 97 & MRN114\_M87150415\_big\_graph\_w\_inv \\ \hline
48 & MRN114\_M87108094\_big\_graph\_w\_inv & 98 & MRN114\_M87150639\_big\_graph\_w\_inv \\ \hline
49 & MRN114\_M87111487\_big\_graph\_w\_inv & 99 & MRN114\_M87151117\_big\_graph\_w\_inv \\ \hline
50 & MRN114\_M87111924\_big\_graph\_w\_inv & 100 & MRN114\_M87151146\_big\_graph\_w\_inv \\ \hline 
\pagebreak
\hline
sample\# & graph & sample\# & graph \\ \hline
101 & MRN114\_M87151453\_big\_graph\_w\_inv & 151 & MRN114\_M87192995\_big\_graph\_w\_inv \\ \hline
102 & MRN114\_M87152844\_big\_graph\_w\_inv & 152 & MRN114\_M87193409\_big\_graph\_w\_inv \\ \hline
103 & MRN114\_M87153569\_big\_graph\_w\_inv & 153 & MRN114\_M87196363\_big\_graph\_w\_inv \\ \hline
104 & MRN114\_M87154559\_big\_graph\_w\_inv & 154 & MRN114\_M87196591\_big\_graph\_w\_inv \\ \hline
105 & MRN114\_M87155496\_big\_graph\_w\_inv & 155 & MRN114\_M87199297\_big\_graph\_w\_inv \\ \hline
106 & MRN114\_M87155949\_big\_graph\_w\_inv & 156 & MRN114\_M87199728\_big\_graph\_w\_inv \\ \hline
107 & MRN114\_M87156106\_big\_graph\_w\_inv & 157 & NKI-TRT\_0021001\_1\_big\_graph\_w\_inv \\ \hline
108 & MRN114\_M87157827\_big\_graph\_w\_inv & 158 & NKI-TRT\_0021001\_2\_big\_graph\_w\_inv \\ \hline
109 & MRN114\_M87158338\_big\_graph\_w\_inv & 159 & NKI-TRT\_0021002\_1\_big\_graph\_w\_inv \\ \hline
110 & MRN114\_M87158534\_big\_graph\_w\_inv & 160 & NKI-TRT\_0021002\_2\_big\_graph\_w\_inv \\ \hline
111 & MRN114\_M87159410\_big\_graph\_w\_inv & 161 & NKI-TRT\_0021006\_1\_big\_graph\_w\_inv \\ \hline
112 & MRN114\_M87159580\_big\_graph\_w\_inv & 162 & NKI-TRT\_0021006\_2\_big\_graph\_w\_inv \\ \hline
113 & MRN114\_M87160332\_big\_graph\_w\_inv & 163 & NKI-TRT\_0021018\_1\_big\_graph\_w\_inv \\ \hline
114 & MRN114\_M87160375\_big\_graph\_w\_inv & 164 & NKI-TRT\_0021018\_2\_big\_graph\_w\_inv \\ \hline
115 & MRN114\_M87161235\_big\_graph\_w\_inv & 165 & NKI-TRT\_0021024\_1\_big\_graph\_w\_inv \\ \hline
116 & MRN114\_M87161902\_big\_graph\_w\_inv & 166 & NKI-TRT\_0021024\_2\_big\_graph\_w\_inv \\ \hline
117 & MRN114\_M87162915\_big\_graph\_w\_inv & 167 & NKI-TRT\_1427581\_2\_big\_graph\_w\_inv \\ \hline
118 & MRN114\_M87164412\_big\_graph\_w\_inv & 168 & NKI-TRT\_1793622\_1\_big\_graph\_w\_inv \\ \hline
119 & MRN114\_M87164886\_big\_graph\_w\_inv & 169 & NKI-TRT\_1793622\_2\_big\_graph\_w\_inv \\ \hline
120 & MRN114\_M87165017\_big\_graph\_w\_inv & 170 & NKI-TRT\_1961098\_1\_big\_graph\_w\_inv \\ \hline
121 & MRN114\_M87165441\_big\_graph\_w\_inv & 171 & NKI-TRT\_1961098\_2\_big\_graph\_w\_inv \\ \hline
122 & MRN114\_M87166115\_big\_graph\_w\_inv & 172 & NKI-TRT\_2475376\_1\_big\_graph\_w\_inv \\ \hline
123 & MRN114\_M87168759\_big\_graph\_w\_inv & 173 & NKI-TRT\_2475376\_2\_big\_graph\_w\_inv \\ \hline
124 & MRN114\_M87174803\_big\_graph\_w\_inv & 174 & NKI-TRT\_2799329\_1\_big\_graph\_w\_inv \\ \hline
125 & MRN114\_M87176019\_big\_graph\_w\_inv & 175 & NKI-TRT\_2799329\_2\_big\_graph\_w\_inv \\ \hline
126 & MRN114\_M87176708\_big\_graph\_w\_inv & 176 & NKI-TRT\_2842950\_1\_big\_graph\_w\_inv \\ \hline
127 & MRN114\_M87178630\_big\_graph\_w\_inv & 177 & NKI-TRT\_2842950\_2\_big\_graph\_w\_inv \\ \hline
128 & MRN114\_M87179511\_big\_graph\_w\_inv & 178 & NKI-TRT\_3201815\_1\_big\_graph\_w\_inv \\ \hline
129 & MRN114\_M87179597\_big\_graph\_w\_inv & 179 & NKI-TRT\_3201815\_2\_big\_graph\_w\_inv \\ \hline
130 & MRN114\_M87179713\_big\_graph\_w\_inv & 180 & NKI-TRT\_3313349\_1\_big\_graph\_w\_inv \\ \hline
131 & MRN114\_M87181205\_big\_graph\_w\_inv & 181 & NKI-TRT\_3313349\_2\_big\_graph\_w\_inv \\ \hline
132 & MRN114\_M87181216\_big\_graph\_w\_inv & 182 & NKI-TRT\_3315657\_1\_big\_graph\_w\_inv \\ \hline
133 & MRN114\_M87182922\_big\_graph\_w\_inv & 183 & NKI-TRT\_3315657\_2\_big\_graph\_w\_inv \\ \hline
134 & MRN114\_M87183189\_big\_graph\_w\_inv & 184 & NKI-TRT\_3795193\_1\_big\_graph\_w\_inv \\ \hline
135 & MRN114\_M87183485\_big\_graph\_w\_inv & 185 & NKI-TRT\_3795193\_2\_big\_graph\_w\_inv \\ \hline
136 & MRN114\_M87184910\_big\_graph\_w\_inv & 186 & NKI-TRT\_3808535\_1\_big\_graph\_w\_inv \\ \hline
137 & MRN114\_M87185000\_big\_graph\_w\_inv & 187 & NKI-TRT\_3808535\_2\_big\_graph\_w\_inv \\ \hline
138 & MRN114\_M87186642\_big\_graph\_w\_inv & 188 & NKI-TRT\_3893245\_2\_big\_graph\_w\_inv \\ \hline
139 & MRN114\_M87187090\_big\_graph\_w\_inv & 189 & NKI-TRT\_4176156\_1\_big\_graph\_w\_inv \\ \hline
140 & MRN114\_M87187750\_big\_graph\_w\_inv & 190 & NKI-TRT\_4176156\_2\_big\_graph\_w\_inv \\ \hline
141 & MRN114\_M87187984\_big\_graph\_w\_inv & 191 & NKI-TRT\_4288245\_1\_big\_graph\_w\_inv \\ \hline
142 & MRN114\_M87188000\_big\_graph\_w\_inv & 192 & NKI-TRT\_4288245\_2\_big\_graph\_w\_inv \\ \hline
143 & MRN114\_M87188762\_big\_graph\_w\_inv & 193 & NKI-TRT\_6471972\_1\_big\_graph\_w\_inv \\ \hline
144 & MRN114\_M87190609\_big\_graph\_w\_inv & 194 & NKI-TRT\_7055197\_1\_big\_graph\_w\_inv \\ \hline
145 & MRN114\_M87190745\_big\_graph\_w\_inv & 195 & NKI-TRT\_7055197\_2\_big\_graph\_w\_inv \\ \hline
146 & MRN114\_M87191087\_big\_graph\_w\_inv & 196 & NKI-TRT\_8574662\_1\_big\_graph\_w\_inv \\ \hline
147 & MRN114\_M87191258\_big\_graph\_w\_inv & 197 & NKI-TRT\_8735778\_1\_big\_graph\_w\_inv \\ \hline
148 & MRN114\_M87192333\_big\_graph\_w\_inv & 198 & NKI-TRT\_8735778\_2\_big\_graph\_w\_inv \\ \hline
149 & MRN114\_M87192557\_big\_graph\_w\_inv & 199 & NKI-TRT\_9630905\_1\_big\_graph\_w\_inv \\ \hline
150 & MRN114\_M87192637\_big\_graph\_w\_inv & 200 & NKI-TRT\_9630905\_2\_big\_graph\_w\_inv \\ \hline
\caption{Sample numbers of brain graphs that we use to identify them in our figures. Graph data is available with identical file names at \cite{s_connectome1}.}
\label{table-samples}
\end{longtable}
\end{center}
\normalsize

\clearpage

\begin{figure}[h!]
\begin{center}
\includegraphics[width=5in]{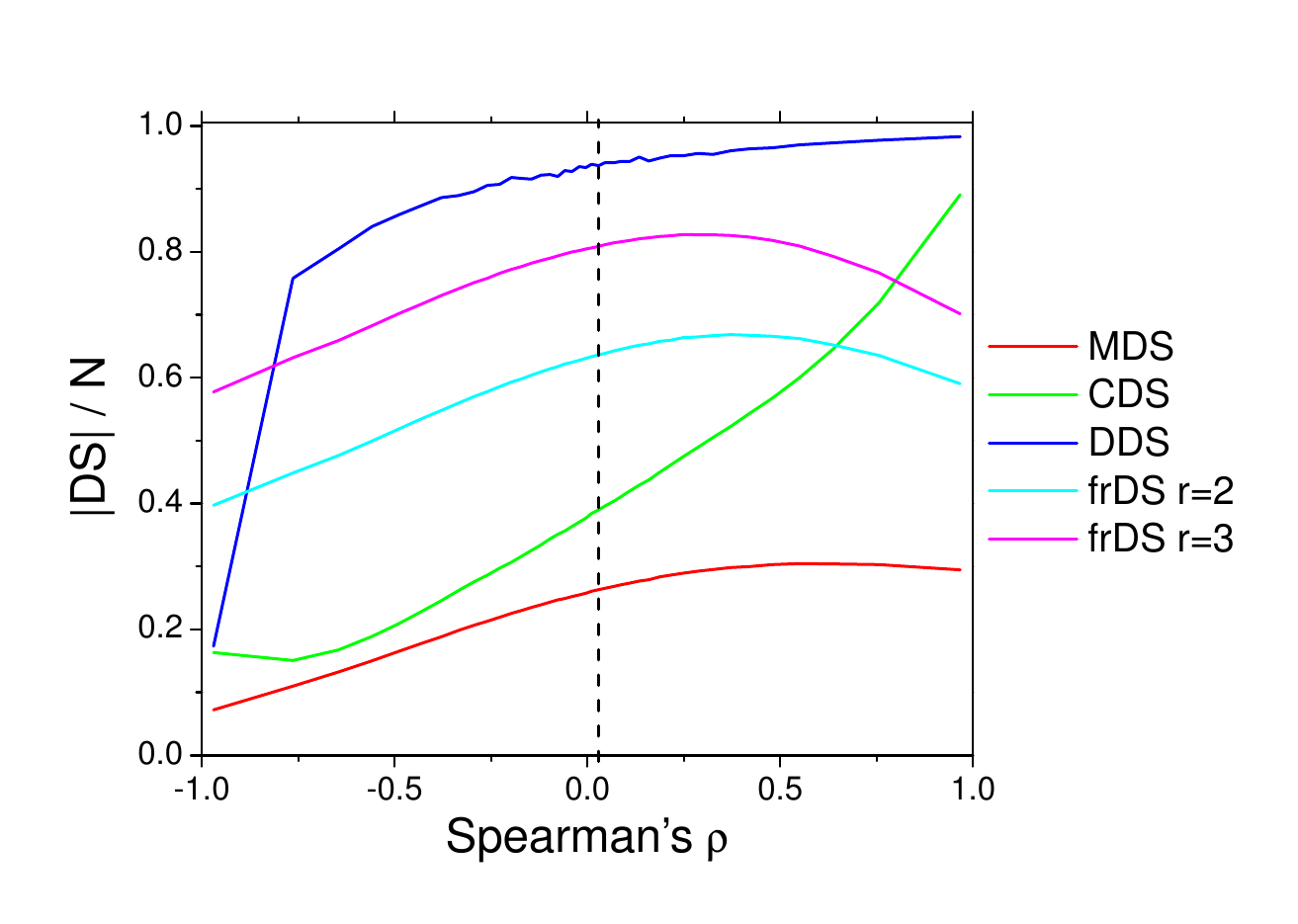}
\caption{Dominating set sizes vs. assortativity in {\bf Gnutella08} graph \cite{s_stanford}, achieved by random (biased) mixing of exges by double-edge swaps. The vertical dashed line indicates the assortativity of the original graph.}
\label{fig-ds-rho-gnutella}
\end{center}
\end{figure}

\begin{figure}[h!]
\begin{center}
\includegraphics[width=5in]{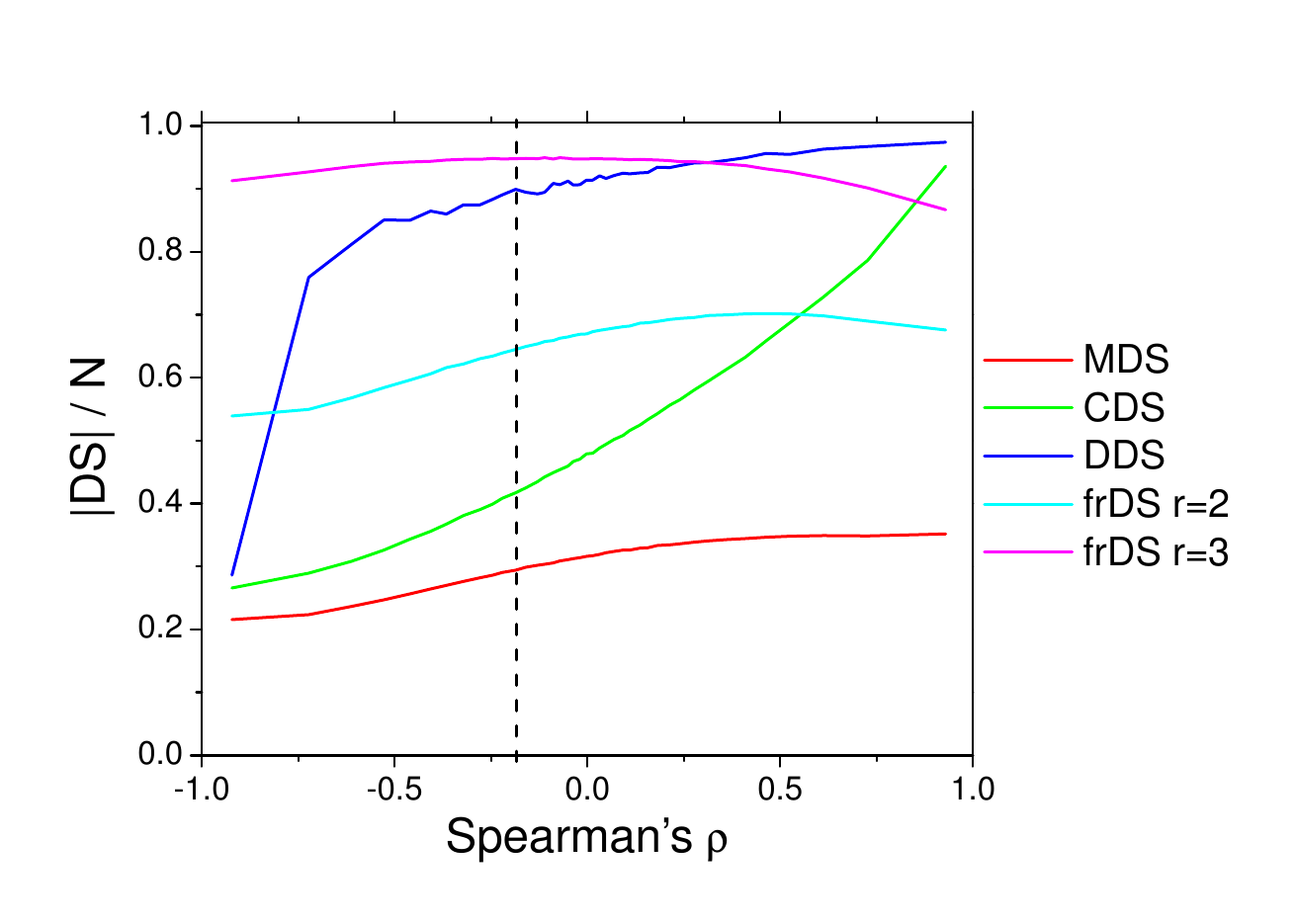}
\caption{Dominating set sizes vs. assortativity in {\bf powergrid} graph \cite{s_powergrid1,s_powergrid2}, achieved by random (biased) mixing of exges by double-edge swaps. The vertical dashed line indicates the assortativity of the original graph.}
\label{fig-ds-rho-pgrid}
\end{center}
\end{figure}

\begin{figure}[h!]
\begin{center}
\includegraphics[width=5in]{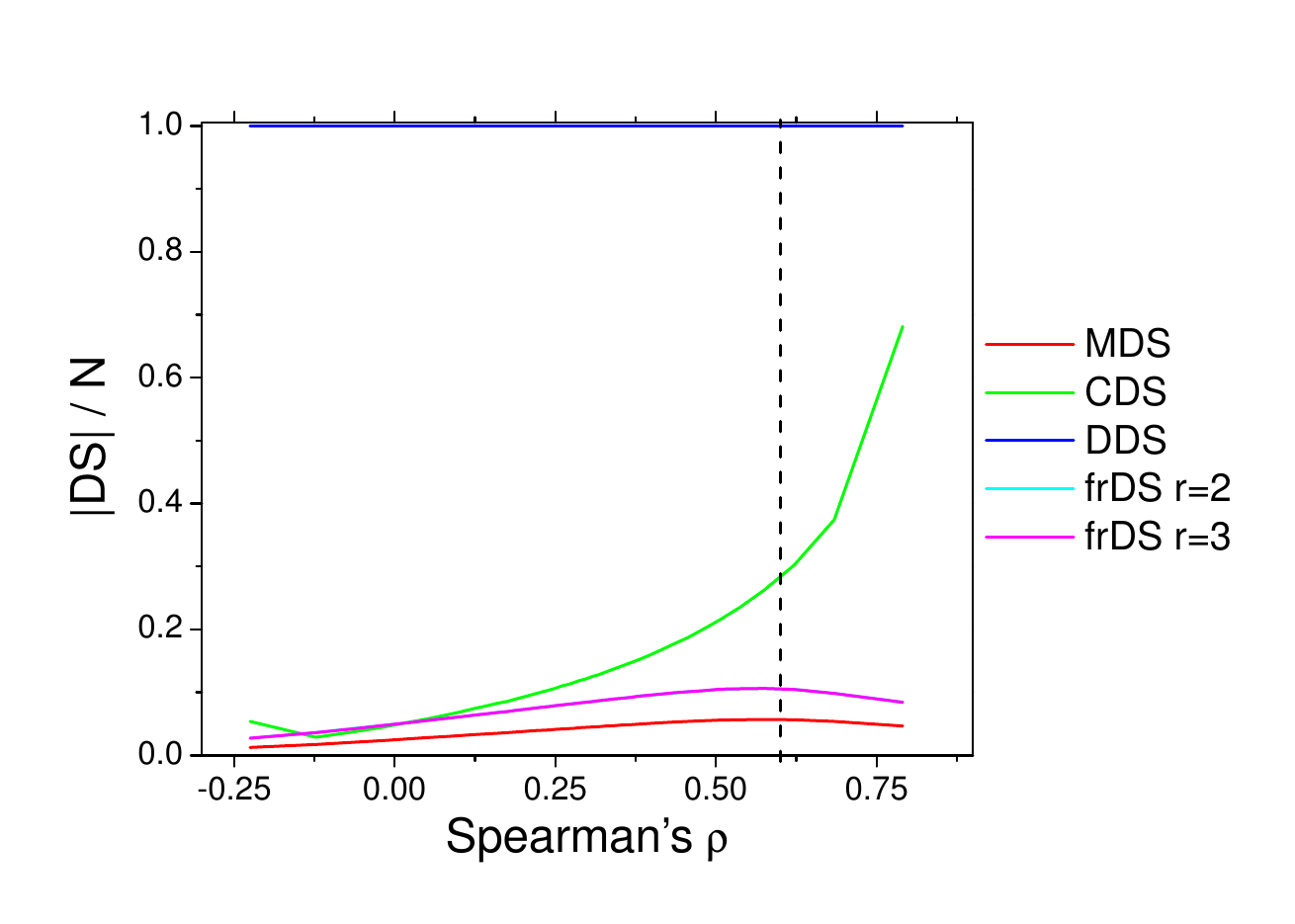}
\caption{Dominating set sizes vs. assortativity in brain graph {\bf KKI-21\_KKI2009-19} \cite{s_connectome1,s_connectome2}, achieved by random (biased) mixing of edges by double-edge swaps. The vertical dashed line indicates the assortativity of the original graph. Note that frDS curves with $r=2$ and $r=3$ overlap.}
\label{fig-ds-rho-brain}
\end{center}
\end{figure}

\end{document}